\documentclass[a4paper,fleqn]{cas-dc}

\usepackage[numbers,sort&compress]{natbib}
\usepackage{amsmath,amsfonts}
\usepackage{algorithmic}
\usepackage{algorithm}
\usepackage{array}
\usepackage[caption=false,font=normalsize,labelfont=sf,textfont=sf]{subfig}
\usepackage{textcomp}
\usepackage{graphicx}
\usepackage{threeparttable}
\usepackage{xcolor}
\usepackage{url}
\usepackage[section]{placeins}
\usepackage{makecell}
\usepackage{stfloats}
\usepackage{xinttools} 
\usepackage{tikz}
\usetikzlibrary{calc}
\usepackage{booktabs}
\usepackage{multirow}

\author[1]{Le Ma}[role=Equal]
\fnmark[1]
\author[2]{Ran Zhang}[role=Equal]
\fnmark[1]
\author[3]{Yikun Han}[role=Equal]
\fnmark[1]
\author[4,5]{Shirui Yu}
\author[2]{Zaitian Wang}
\author[2]{Zhiyuan Ning}
\author[6]{Jinghan Zhang}
\author[2]{Ping Xu}
\author[2]{Pengjiang Li}
\author[7]{Ziyue Qiao}
\author[8]{Wei Ju}
\author[9]{Chong Chen}
\author[10]{Dongjie Wang}
\author[6]{Kunpeng Liu}
\author[11]{Pengyang Wang}
\author[2]{Pengfei Wang}
\author[12]{Yanjie Fu}
\author[4,5]{Chunjiang Liu}[role=Corresponding]
\cormark[1]
\author[2]{Yuanchun Zhou}
\author[13]{Chang-Tien Lu}

\affiliation[1]{organization={Sichuan University},
            addressline={Department of Information Resources Management, School of Public Administration}, 
            city={Chengdu},
            postcode={610041}, 
            country={China}}

\affiliation[2]{organization={Computer Network Information Center, Chinese Academy of Sciences},
            addressline={University of Chinese Academy of Sciences}, 
            city={Beijing},
            postcode={100190}, 
            country={China}}
            
\affiliation[3]{organization={University of Illinois Urbana-Champaign},
            addressline={School of Information Sciences}, 
            city={Champaign},
            postcode={61820}, 
            country={United States}}
            
\affiliation[4]{organization={National Science Library (Chengdu), Chinese Academy of Sciences},
            addressline={Department of Information Resources Management, School of Economics and Management}, 
            city={Chengdu},
            postcode={610041}, 
            country={China}}

\affiliation[5]{organization={University of Chinese Academy of Sciences},
            addressline={Department of Information Resources Management, School of Economics and Management}, 
            city={Beijing},
            postcode={101408}, 
            country={China}}

\affiliation[6]{organization={Portland State University},
            addressline={Department of Computer Science}, 
            city={Portland},
            postcode={101408}, 
            country={United States}}

\affiliation[7]{organization={Great Bay University},
            addressline={School of Computing and Information Technology}, 
            city={Dongguan},
            postcode={523000}, 
            country={China}}

\affiliation[8]{organization={Sichuan University},
            addressline={College of Computer Science}, 
            city={Chengdu},
            postcode={610041}, 
            country={China}}

\affiliation[9]{organization={Terminus Group},
            addressline={the Future City Lab}, 
            city={Beijing},
            postcode={100032}, 
            country={China}}

\affiliation[10]{organization={University of Kansas},
            addressline={Department of Electrical Engineering and Computer Science}, 
            city={Lawrence},
            postcode={66045}, 
            country={United States}}            

\affiliation[11]{organization={University of Macau},
            addressline={Department of Computer and Information Science, The State Key Laboratory of Internet of Things for Smart City}, 
            city={Macau},
            postcode={999078}, 
            country={China}} 

\affiliation[12]{organization={Arizona State University},
            addressline={}, 
            city={Tempe},
            postcode={85287}, 
            country={United States}} 

\affiliation[13]{organization={Virginia Tech},
            addressline={}, 
            city={Blacksburg},
            postcode={24061}, 
            country={United States}}
            
\title[mode=title]{A Comprehensive Survey on Vector Database: Storage and Retrieval Technique, Challenge}
\shorttitle{Survey on Vector Database}
\shortauthors{Ma et al.}

\tnotemark[1,2]
\tnotetext[1]{This research was supported by the National Natural Science Foundation of China.}
\tnotetext[2]{This work was partially supported by the Chinese Academy of Sciences.}

\fntext[1]{These authors contributed equally to this work.}
\cortext[1]{Corresponding author.}
\ead{liucj@clas.ac.cn}

\credit{Conceptualization, Methodology, Investigation, Writing - Original Draft, Writing - Review \& Editing}

\begin{document}


\begin{abstract}
[Objective] As high‑dimensional vector data increasingly surpasses the processing capabilities of traditional database management systems, Vector Databases (VDBs) have emerged and become tightly integrated with large language models, being widely applied in modern artificial intelligence systems. However, existing research has primarily focused on underlying technologies such as approximate nearest neighbor search, with relatively few studies providing a systematic architectural‑level review of VDBs or analyzing how these core technologies collectively support the overall capacity of VDBs. This survey aims to offer a comprehensive overview of the core designs and algorithms of VDBs, establishing a holistic understanding of this rapidly evolving field. [Methods] First, we systematically review the key technologies and design principles of VDBs from the two core dimensions of storage and retrieval, tracing their technological evolution. Next, we conduct an in‑depth comparison of several mainstream VDB architectures, summarizing their strengths, limitations, and typical application scenarios. Finally, we explore emerging directions for integrating VDBs with large language models, including open research challenges and trends such as novel indexing strategies. [Conclusions]This survey serves as a systematic reference guide for researchers and practitioners, helping readers quickly grasp the technological landscape and development trends in the field of vector databases, and promoting further innovation in both theoretical and applied aspects.

\end{abstract}





\vspace{-3mm}
\maketitle
\section{Introduction} 

{Vectors}, particularly those in high-dimensional spaces, are mathematical representations of data, encoding the semantic and contextual information of entities such as text, images, audio, and video~\cite{cao2024knowledge, 10.1145/3150226}. 
These vectors are generally generated through some related machine learning models, and the generated vectors are usually high-dimensional and can be used for similarity comparison. The step of converting original unstructured data into vectors is the foundation of many artificial intelligence (AI) applications (including large language models (LLMs)~\cite{zhao2023survey}, question-answering systems~\cite{allam2012question, biancofiore2024interactive}, image recognition~\cite{7445232}, recommendation systems~\cite{zhao2024recommender, liu2024multimodal}, etc.). However, in terms of managing and retrieving high-dimensional vector data, traditional databases designed for handling structured data are often inadequate. Vector databases (VDBs), on the other hand, provide a specialized solution to these challenges. 

\begin{figure*}[thbp] 
    \centering
    \vspace{-3mm}
    \includegraphics[width=0.8\textwidth]{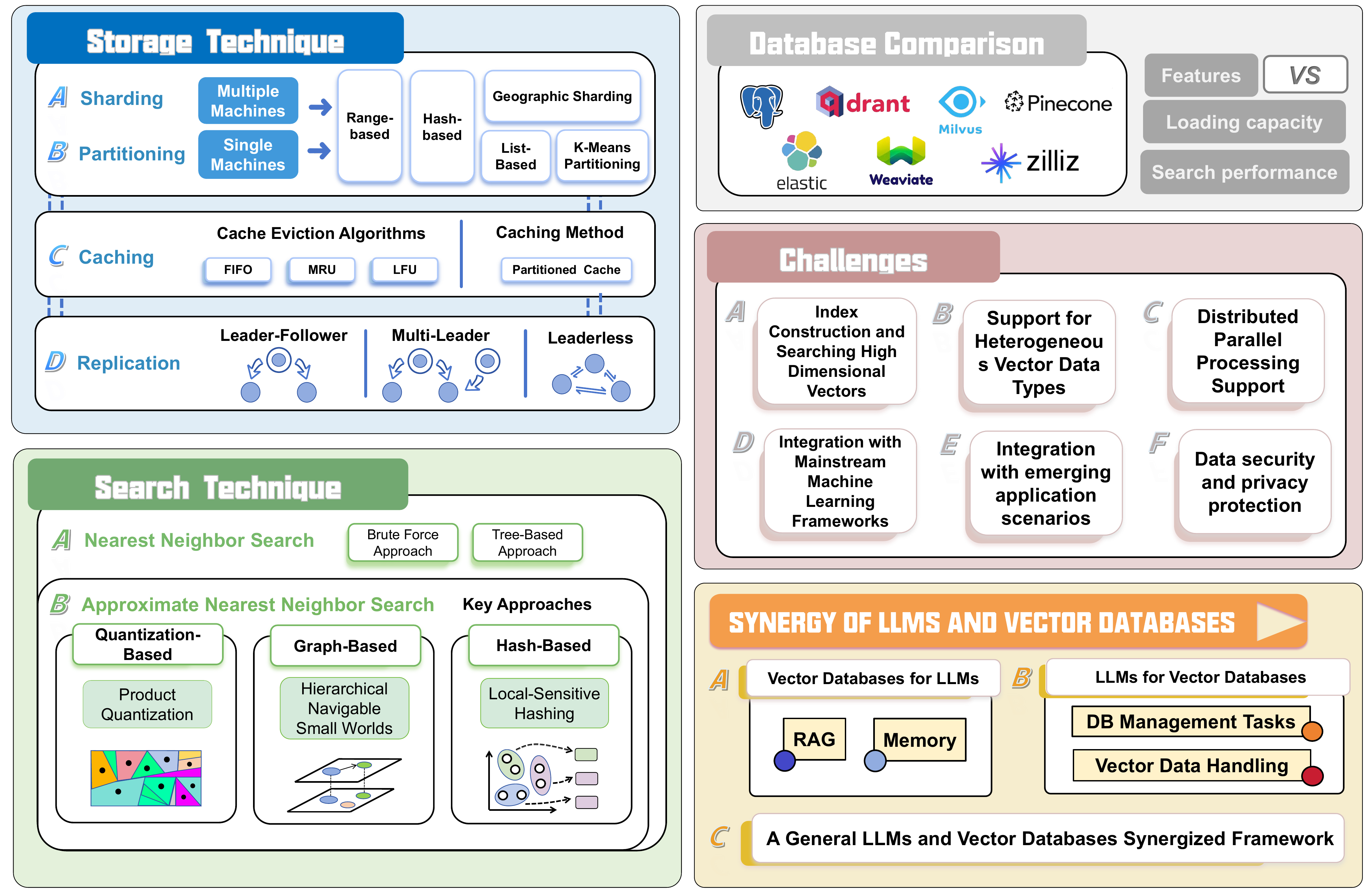} 
    \vspace{-2mm}
    \caption{Framework overview of this survey structure covering Storage Techniques, Search Techniques, Database Comparison, Challenges, and the Synergy of Large Language Models (LLMs) with VDBs. Each section represents a fundamental facet of the operation and integration of modern VDBs within advanced AI technologies.}
    \vspace{-3mm}
    \label{fig:overview}
\end{figure*}

VDBs are tools specifically designed to efficiently store and manage high-dimensional vectors. Specifically, VDBs store information as high-dimensional vectors, which are mathematical representations of data features or attributes~\cite{10.1145/3382080}. Depending on the complexity and granularity of the underlying data, the dimensions of these high-dimensional vectors usually range from dozens to thousands. Unlike traditional relational databases, VDBs provide efficient mechanisms for large-scale storage, management, and search of high-dimensional vectors~\cite{wang2024efficient,kraska2018case, xie2023bri}. These mechanisms bring various efficient functions to VDBs, such as supporting semantic similarity search, efficiently managing large-scale data, and providing low-latency responses. These functions make VDBs increasingly integrated into AI-based applications. 
VDBs have two core functions: vector storage and vector retrieval. The vector storage function relies on techniques such as quantization, compression, and distributed storage mechanisms to improve efficiency and scalability. The retrieval function of VDBs relies on specialized indexing techniques, including tree-based methods, hashing methods~\cite{zhao2023towards}, graph-based models, and quantization-based techniques~\cite{wang2023graph}. These indexing techniques optimize high-dimensional similarity search by reducing computational cost and improving search performance.  In addition, hardware acceleration and cloud-based technologies have further enhanced the capabilities of VDBs, making them suitable for large-scale and real-time applications~\cite{li2019approximate, karthik2024bang, jegou2022faiss}.

Consequently, VDBs offer three key benefits over traditional databases: \textit{{\textbf{(1) VDBs can retrieve vectors accurately and efficiently.}}}VDBs' primary purpose is to retrieve pertinent vectors based on vector similarity (distance); this function is also at the heart of applications like recommendation systems, computer vision, and natural language processing (NLP)~\cite{mohoney2023high, pan2024vector}. In contrast, traditional databases can only query data based on exact matching or predefined conditions, and this kind of query method is relatively slow and often does not consider the semantic information of the data itself. 
\textit{{\textbf{(2) VDBs can be used to store and query complex and unstructured data.}}} Text, images, audio, video, and other types of data can all be stored and searched with great granularity and complexity using VDBs. However, traditional databases, such as relational databases, are difficult to store this kind of data information well~\cite{pan2024survey}.
\textit{{\textbf{(3) VDBs are very scalable and have real-time data processing capabilities.}}} Large volumes of real-time datasets and unstructured data are difficult for traditional databases to handle efficiently~\cite{rao2019big}. 

However, due to their ability to process vector data on a large scale and in real time, VDBs are crucial for modern data science and artificial intelligence~\cite{wang2024starling}. Through the use of technologies like caching, replication, partitioning, and sharding~\cite{su2024vexless}, VDBs can optimize resource utilization and divide workloads among multiple machines or clusters. However, when dealing with big data, traditional databases might experience concurrency conflicts, latency problems, or scalability bottlenecks~\cite{pan2024vector}.

Recent surveys on VDBs primarily cover fundamental concepts and practical applications of VDBs and vector database management systems. 
Some studies~\cite{xie2023bri,taipalus2024vector,pan2024vector} focus on the workflow and technical challenges of VDBs, including key aspects such as query processing, optimization, and execution techniques. 
And some works~\cite{joshiintroduction} explore the critical role of VDBs in modern generative AI applications and provide an outlook on the future of VDBs. 
While these studies have their respective focuses, they do not provide a comprehensive survey of the overall storage and search technologies in VDBs, nor deliver a thorough analysis comparing the capabilities of existing VDBs.
Furthermore, there is limited exploration of how these systems can integrate with rapidly advancing AI technologies, such as large language models (LLMs), to support modern data-intensive applications. 

\begin{figure*}[htp] 
    \centering
    \includegraphics[width=0.8\textwidth]{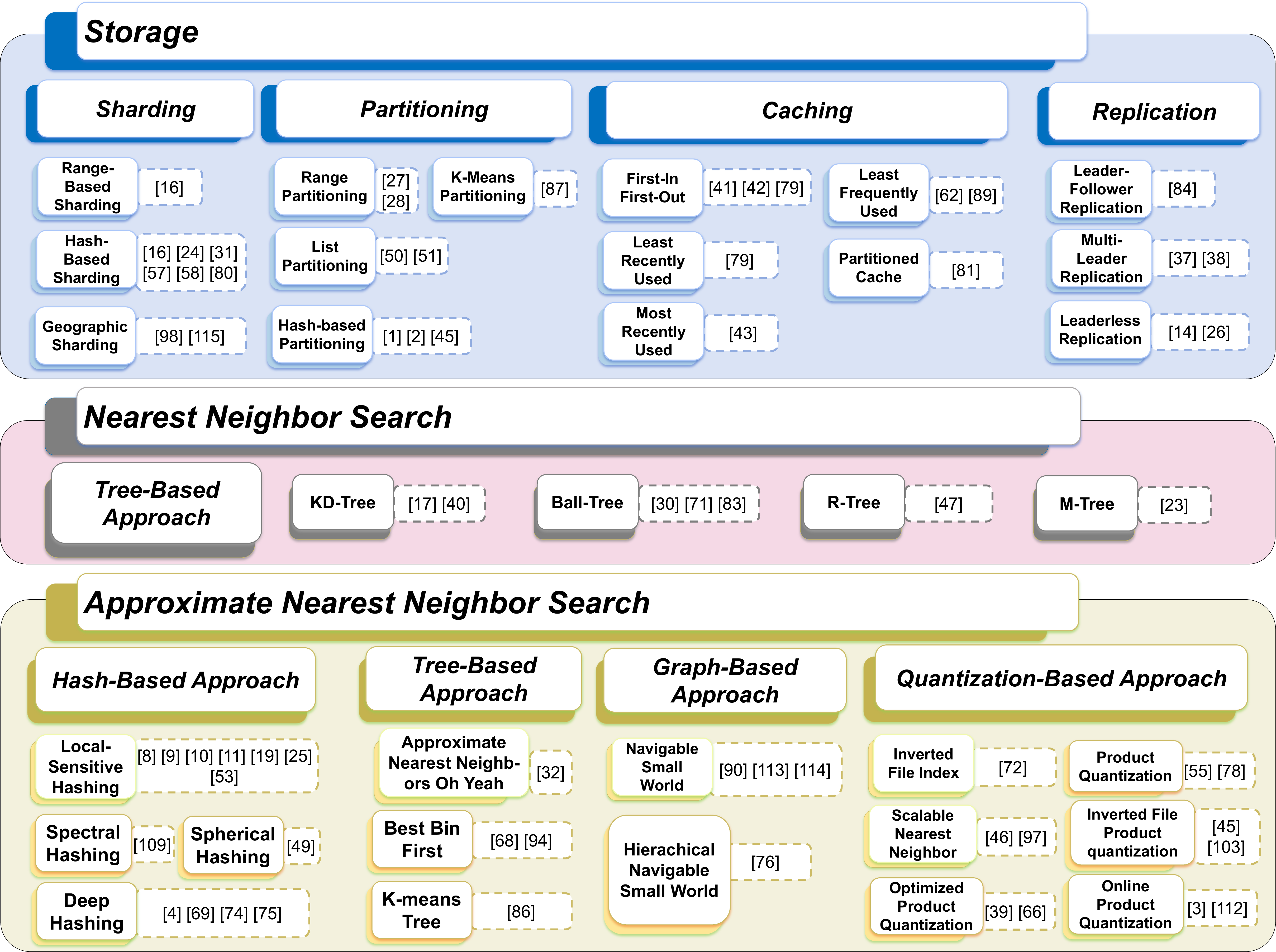} 
    \caption{Taxonomy of VDB Storage and Search Technologies.}
    \label{fig:technique framework}
\end{figure*}

This research gap emphasizes the need for a thorough survey that aims to integrate the existing body of knowledge regarding VDBs and identify the main research issues that require immediate attention. The following are the main contributions our survey makes to address this issue:
\begin{itemize}
    \item We organized and reviewed the storage and retrieval technologies in VDBs in a methodical manner.
    \item To highlight the advantages and disadvantages of each open-source VDB, we conducted thorough performance testing and comparison.
    \item We went into great detail about the primary issues that VDBs are currently facing and how they integrate with Large Language Models (LLMs). 
\end{itemize} 
This paper comprehensively summarizes the technologies related to VDBs and systematically tests the performance of existing open-source VDBs. It also provides an outlook on the challenges that VDBs will face in the future. Through the summary of this paper, researchers can deepen their understanding of the field of VDBs. Figure~\ref{fig:overview}  shows the overall framework of the paper, and we also construct a classification system of storage and search technologies for VDBs, as shown in Figure~\ref{fig:technique framework}.

\section{Storage}
Efficient data management strategies are essential for the performance and scalability of VDBs. 
This section explores four key techniques: sharding and partitioning for data distribution, caching for reducing query latency, and replication for ensuring availability and fault tolerance. 

\vspace{-3mm}
\subsection{Sharding}
Sharding spreads a VDB across several computers or clusters, called shards, by routing records using a rule such as a hash or a key range. By cutting the dataset into smaller pieces, sharding helps the system scale, shares work evenly, and adds resilience when a single node fails. Different sharding methods target specific goals—such as easy expansion, hotspot reduction, fair size balancing, and quicker queries—so designers pick the one that fits their load. Here we review three popular schemes: range-based, hash-based, and geographic sharding.

\textbf{Range-Based Sharding.} 
Range-based sharding is the simplest approach; it splits sorted keys into clean, non-overlapping intervals and sends each interval to its own shard ~\cite{oracle2021shard}. 
For example, users can shard a VDB by dividing the ID column of the vector data into different ranges, such as 0-999, 1000-1999, 2000-2999, and so on, with each range assigned to a specific shard.  
Each range corresponds to a shard. This way, users can query the vector data more efficiently by specifying the shard name or range. Range-based sharding provides efficient query performance and is simple to use for range-based queries, such as retrieving all vectors within a specified time frame or ID range. However, if key values are not distributed uniformly, this approach frequently results in data skew and uneven load distribution. Use cases like time-series analytics or sequential ID-based queries, where data access patterns are predictable, are best suited for range-based sharding.

\textbf{Hash-Based Sharding}. Another common sharding method in VDBs is hash-based sharding~\cite{costa2015sharding,author2024ebook,oracle2021shard}, which assigns vector data into different shards based on the hash value of a key column or a set of columns. 
For example, users can shard a VDB by applying a hash function to the ID column of the vector data. 
This lets users spread vector data evenly across the shards and avoid hotspots. But traditional hash-based methods, like modulo hashing, cause problems when scaling a cluster. When users add or remove shards, a large amount of data must be moved. To fix this, many distributed VDBs use consistent hashing. Unlike regular hashing, consistent hashing puts both data and nodes on a ring-shaped hash space. When a node is added or removed, only a small part of the data (O(1/N)) needs to move. This helps reduce the amount of data that must be changed during scaling. It also keeps the data balanced across the system. Also, virtual nodes (copies of physical nodes on the hash ring) help with load balancing. They spread the traffic more evenly and reduce hotspots. Although consistent hashing introduces slight computational complexity, its advantages in dynamic environments make it a preferred choice for modern distributed systems~\cite{mirrokni2018consistent,karger1999web,karger1997consistent}.

\textbf{Geographic Sharding.} 
This method puts data into shards based on location, such as user region ~\cite{taft2020cockroachdb,zhang2013transaction}. A VDB can split data using a "region" field and send it to different zones like "North America" or "Europe." Geo-sharding helps with apps that need low delay, like recommendation systems. Storing data close to users lowers network delay. It also helps follow local data regulations. 

\vspace{-3mm}
\subsection{Partitioning}

Partitioning means splitting data in one database into smaller parts based on simple rules like range, list, or K-means. All the data stays in the same physical system. This method helps make queries faster. For example, users can split a VDB by color values like red, yellow, and blue. Each part holds vector data with a certain value in the “color” column. Queries can then target specific partitions, avoiding unnecessary scans of unrelated data. While sharding is used to distribute data across multiple machines, it helps systems scale horizontally. Partitioning, on the other hand, is used to organize data within a single machine to improve local access and performance. Using both together helps the VDBs handle large data and fast queries—sharding spreads the data across many machines, and partitioning makes data access faster inside each shard. Partitioning strategies change based on the type of data and what the application needs. This section looks at four common methods. These are range partitioning, list partitioning, k-means partitioning, and hash-based partitioning. Each method has its own use case and works best under different conditions.

\textbf{Range-based Partitioning.} 
Range-based partitioning is a method widely used in VDBs, where data is divided into non-overlapping key ranges to form partitions~\cite{dewitt1992parallel, dewitt1990hybrid}. 
Each range corresponds to a specific subset of data based on a sorted key (e.g., timestamps, numeric IDs). 
Similar to the strength of range-based sharding, range-based partitioning is particularly efficient for range-based queries, as it allows the system to target specific partitions directly. 
For example, users can partition a VDB by date ranges, such as monthly or quarterly. 
This way, users can query the vector data more efficiently by specifying the partition name or range. 

\textbf{List-based Partitioning.} Another way that partitioning works in VDB is by
using a list partitioning method, which assigns vector data to
different partitions based on their value lists of a key column
or a set of columns~\cite{hobbs2011oracle,hotka2002oracle9i}. For example, users can partition a vector
database by color values, such as red, yellow, and blue. Each
partition contains vector data that has a given value of the
color column. This way, users can query the vector data more
easily by specifying the partition name or lists.

\textbf{K-Means Partitioning.} 
A third partitioning method for VDBs is k-means partitioning~\cite{pan2024survey}, which divides vector data into a predetermined number (k) of clusters. 
Each cluster represents a partition, with vectors within the same cluster being similar to each other and vectors between clusters differing significantly. 
Similar vectors are placed in the same partition, which improves query efficiency. 
However, for large-scale datasets, the computational cost of k-means clustering can be high, especially when frequent updates necessitate re-clustering. 

\textbf{Hash-based Partitioning.} 
Alternatively, some VDBs adopt hash-based partitioning, such as consistent or uniform hashing~\cite{weaviate,vespa,  guo2022manu}. 
Specifically, the hashing partitioning strategy uses a hash function to map data to different partitions. 
The hash value of each data point determines which partition it belongs to. 
In VDBs, the hash value is typically calculated based on certain features of the vector (for example, values of specific dimensions or the entire vector). 
The hash function can evenly distribute data across partitions, preventing any single partition from storing too much data. 
However, when the data distribution is uneven, it may lead to some partitions becoming overloaded. 
Additionally, when the node number changes, hash-based partitioning may require the redistribution of the large dataset, which can incur significant overhead. 

\vspace{-3mm}
\subsection{Caching}
Caching plays a central role in optimizing the efficiency of data-intensive systems. In the context of VDBs, caching helps reduce latency and improve retrieval speed by temporarily storing vectors that are frequently accessed or recently used. While traditional databases often rely on in-memory key-value stores like Redis to cache structured data using predefined query keys, this approach does not directly translate to the vector space. The reason lies in the nature of the data: vectors are high-dimensional and continuous, making exact matches across queries rare.
As VDBs increasingly serve workloads such as similarity search or embedding retrieval, the need for tailored caching strategies becomes clear. General caching algorithms—originally designed for web content or structured data—must be revisited in light of the unique access patterns and data characteristics found in vector systems. In this section, we discuss four widely used caching strategies—first-in first-out (FIFO), least recently used (LRU), most recently used (MRU), and least frequently used (LFU)—as well as partitioned caching in VDB design.
 
\textbf{First-In First-Out (FIFO)}. 
FIFO is one of the simplest cache eviction strategies. It works by removing the oldest items in the cache once space runs out~\cite{grund2009abstract,grund2010precise}. Internally, it maintains a queue where new items are added to the back and removals occur from the front~\cite{mattson1970evaluation}. FIFO performs well in environments where data arrives in regular intervals and older entries quickly lose relevance—such as time-series data from IoT sensors. Its appeal lies in its constant-time performance and ease of implementation. However, because it ignores how often items are accessed, it may evict frequently used vectors prematurely.

\textbf{Least Recently Used (LRU)}. 
The LRU algorithm prioritizes keeping recently accessed items in memory. When the cache fills up, it evicts the vector that hasn't been used for the longest time, thereby improving the likelihood of cache hits~\cite{mattson1970evaluation}. This strategy aligns with many real-world patterns—such as search engines or recommendation systems—where recent queries are likely to be repeated. Systems like Redis have incorporated LRU for managing in-memory vectors. That said, LRU can struggle in workloads where access patterns are periodic or long-tailed, since it doesn't track long-term popularity.

\textbf{Most Recently Used (MRU)}. 
MRU takes the opposite stance: it evicts the most recently accessed item~\cite{gu2011theory}. This approach can be useful in scenarios where users rarely re-access the same vector right after retrieval, such as in certain data processing pipelines or streaming tasks. However, in most applications where temporal locality is strong, MRU tends to underperform, as it may discard items that will be queried again soon.

\textbf{Least Frequently Used (LFU)}. 
The LFU algorithm is a frequency-based cache eviction mechanism that determines removal priority by continuously tracking the access count of each vector data item. 
The LFU algorithm maintains a frequency table and evicts the least frequently accessed items when the cache is full. 
A typical implementation requires maintaining an access counter for each cached item, often using a min-heap data structure to efficiently identify the lowest-frequency items with a time complexity of O(log n)~\cite{lee1999existence, podlipnig2003survey}. 
However, the LFU algorithm can be susceptible to cache pollution, as items frequently accessed in the past may linger in the cache even when they are no longer needed. 
Thus, LFU is suitable for applications with stable and long-lived spot data patterns (e.g., popular product recommendations, high-frequency user profile queries), while its effectiveness diminishes in environments with rapidly evolving access distributions.  

\textbf{Partitioned Cache}.
Partitioned caching is a common approach in VDBs, wherein vector data are divided into multiple partitions based on specific criteria~\cite{mittal2017survey}, such as geographic location, category, or access frequency.  
Each partition can have its own size and eviction policy. This method offers greater flexibility and better aligns caching behavior with application needs. For example, geographic information systems (GIS) often partition vector data by region to support fast and localized map rendering. When configured well, partitioned caches can improve hit rates while preventing a few high-frequency. 

\vspace{-3mm}
\subsection{Replication}
Replication means making several copies of vector data and placing them on different nodes or clusters. This helps boost the availability, reliability, and overall performance of a VDB. In this section, we look at three widely used replication strategies: leader-follower, multi-leader, and leaderless replication.

\textbf{Leader-Follower Replication}. Leader-Follower replication designates one node as the leader and the others as the followers and allows only the leader to accept write requests and propagate them to the followers~\cite{ongaro2014search}. Leader-follower replication can ensure strong consistency and simplify the conflict resolution of VDB. However, it may also introduce availability issues and require failover mechanisms to handle leader failures.

\textbf{Multi-Leader Replication}. 
Multi-Leader replication extends the traditional leader-follower model by designating multiple nodes as leaders, each capable of independently accepting and processing write requests~\cite{garcia1985assign,garmany2003oracle}. 
In this architecture, all leader nodes can concurrently handle write operations and asynchronously propagate changes to other nodes in the system. 

\textbf{Leaderless Replication}. Leaderless replication does not distinguish between leader and follower nodes and allows any node to accept write and read requests~\cite{decandia2007dynamo,bailis2014quantifying}. Leaderless replication can avoid single points of failure and improve the scalability and reliability of VDB. However, it may also introduce consistency issues and require coordination mechanisms to resolve conflicts.

\vspace{-3mm}
\section{Search}

VDBs are designed to facilitate efficient similarity search over high-dimensional vector data, an essential operation in many AI and machine learning applications. 
This similarity search is typically implemented through nearest neighbor search algorithms, which can be further divided into exact nearest neighbor search (NNS) and approximate nearest neighbor search (ANNS) methods. 

NNS is the optimization problem of finding the point in a given set that is closest (or most similar) to a given point. 
Closeness is typically expressed in terms of a dissimilarity function: the less similar the objects, the larger the function values. 
For example, users can use NNS to find images that are similar to a given image based on their visual content and style, or documents that are similar to a given document based on their topic and sentiment. 
ANNS is a variation of NNS that allows for some error or approximation in the search results. ANNS can trade off accuracy for speed and space efficiency, which can be useful for large-scale and high-dimensional data. For example, users can use ANNS to find products that are similar to a given product based on their features and ratings, or users that are similar to a given user based on their preferences and behaviors.

 NNS algorithms tend to use more exact or deterministic methods, such as partitioning the space into regions by splitting along one dimension (k-d tree) or enclosing groups of points in hyperspheres (ball tree), and visiting only the regions that may contain the nearest neighbor based on some distance bounds or criteria. ANNS algorithms tend to use more probabilistic or heuristic methods, such as mapping similar points to the same or nearby buckets with high probability (locality-sensitive hashing), visiting the regions in order of their distance to the query point and stopping after a fixed number of regions or points (best bin first), or following the edges that lead to closer points in a graph with different levels of coarseness (hierarchical navigable small world).

In fact, a data structure or algorithm that supports NNS can also be applied to ANNS, and for ease of categorization, such methods are included under the section on NNS. And in recent years, several new algorithms for high-dimensional vector NNS have emerged~\cite{gao2023high, manohar2024parlayann, fu2021high, wang2023dumpy, zhao2023towards}. 
Although these algorithms have not yet been widely adopted by VDBs, they hold significant potential for future applications. 

\vspace{-3mm}
\subsection{Nearest Neighbor Search}

\subsubsection{Brute Force Approach}
A brute force algorithm for the NNS problem scans all points in the dataset,  computing their distances to the query point and tracking the closest one. 
This algorithm guarantees to find the true nearest neighbor for any query point, but it has a high computational cost. 
The time complexity of a brute force algorithm for the NNS problem is $O(n)$, where n is the size of the dataset. 
The space complexity is $O(1)$, since no extra space is needed.

\subsubsection{Tree-Based Approach}
Four tree-based methods will be presented here, namely k-dimensional tree (KD-Tree), Ball-Tree, R-Tree, and M-Tree.

\textbf{KD-Tree}~\cite{bentley1975multidimensional}. It is a technique for organizing points in a k-dimensional space, where k is usually a very big number. It works by building a binary tree in which every node is a k-dimensional point. Every non-leaf node in the tree acts as a splitting hyperplane that divides the space into two parts, known as half-spaces. The splitting hyperplane is perpendicular to the chosen axis, which is associated with one of the k dimensions. The splitting value is usually the median or the mean of the points along that dimension.

The algorithm maintains a priority queue of nodes to visit, sorted by their distance to the query point. At each step, the algorithm pops the node with the smallest distance from the queue and checks if it is a leaf node or an internal node. If it is a leaf node, the algorithm compares the distance between the query point and the data point stored in the node and updates the current best distance and nearest neighbor if necessary. If it is an internal node, the algorithm pushes its left and right children to the queue, with their distances computed as follows:

\vspace{-4mm}
\begin{equation}
\vspace{-3mm}
\small
d_{L}(q, N) =
\begin{cases}
0 & \text{if } q_{\mathit{N.axis}} \leq N.\mathit{value} \\
(q_{\mathit{N.axis}} - N.\mathit{value})^2 & \text{if } q_{\mathit{N.axis}} > N.\mathit{value}
\end{cases}
\label{eq:kdtree-left}
\end{equation}

\begin{equation}
\vspace{-3mm}
\small
d_{R}(q, N) =
\begin{cases}
0 & \text{if } q_{\mathit{N.axis}} \geq N.\mathit{value} \\
(N.\mathit{value} - q_{\mathit{N.axis}})^2 & \text{if } q_{\mathit{N.axis}} < N.\mathit{value}
\end{cases}
\label{eq:kdtree-right}
\end{equation}
where $q$ is the query point, $N$ is the internal node, $N.axis$ is the splitting axis of $N$, and $N.value$ is the splitting value of $N$. The algorithm repeats this process until the queue is empty or a termination condition is met.

The advantage of KD-tree is that it is conceptually simpler and often easier to implement than some of the other tree structures.
The performance of KD-tree depends on several factors, such as the dimensionality of the space, the number of points, and the distribution of the points. There are also some challenges and extensions of KD-tree, such as dealing with the curse of dimensionality when the dimensionality is high, introducing randomness in the splitting process to improve robustness, or using multiple trees to increase recall. This is a variation of KD-tree named randomized KD-tree that introduces some randomness in the splitting process, which can improve the performance of KD-tree by reducing its sensitivity to noise and outliers~\cite{ghojogh2018treebased}.

\textbf{Ball-Tree}~\cite{omohundro1989five,liu2006new,dolatshah_ball-tree_2015}. It is a technique for finding the nearest neighbors of a given vector in a large collection of vectors. It works by building a ball-tree, which is a binary tree that partitions the data points into balls, i.e., hyperspheres that contain a subset of the points. Each node of the ball-tree defines the smallest ball that contains all the points in its subtree. The algorithm then searches for the closest ball to the query point and then searches within the closest ball to find the closest point to the query point.

To query for the nearest neighbor of a given point, the ball tree algorithm uses a priority queue to store the nodes to be visited, sorted by their distance to the query point. The algorithm starts from the root node and pushes its two children to the queue. Then, it pops the node with the smallest distance from the queue and checks if it is a leaf node or an internal node. If it is a leaf node, it computes the distance between the query point and each data point in the node and updates the current best distance and nearest neighbor if necessary. If it is an internal node, it pushes its two children to the queue, with their distances computed as follows:
\vspace{-1mm}
\begin{equation}
\vspace{-1mm}
\small
\begin{array}{l}
\label{eq:Ball-tree}
d_{L}(q, N)=\max (0,  { N.value }-\| q- { N.center } \|) \\
d_{R}(q, N)=\max (0, \| q- { N.center } \|-N . v a l u e)
\end{array}
\end{equation}
where $q$ is the query point, $N$ is the internal node, $N.center$ is the center of the ball associated with $N$, and $N.value$ is the radius of the ball associated with $N$. The algorithm repeats this process until the queue is empty or a termination condition is met.

The advantage of ball-tree is that it can perform well in high-dimensional spaces, as it can avoid the curse of dimensionality that affects other methods , such as KD-tree. 
The performance of ball-tree depends on several factors, such as the dimensionality of the data, the number of balls per node, and the distance approximation method used. There are also some challenges and extensions of ball-tree search, such as dealing with noisy and outlier data, choosing a good splitting dimension and value for each node, or using multiple trees to increase recall.

\textbf{R-Tree}~\cite{guttman1984r}. It is a technique for finding the nearest neighbors of a given vector in a large collection of vectors. It works by building an R-tree, which is a tree data structure that partitions the data points into rectangles, i.e., hyperrectangles that contain a subset of the points. Each node of the R-tree defines the smallest rectangle that contains all the points in its subtree. The algorithm then searches for the closest rectangle to the query point and then searches within the closest rectangle to find the closest point to the query point.

The R-tree algorithm uses the concept of minimum bounding rectangle (MBR) to represent the spatial objects in the tree. The MBR of a set of points is the smallest rectangle that contains all the points. The formula for computing the MBR of a set of points  $P$  is:
\vspace{-4mm}

\begin{equation}
\vspace{-2mm}
\small
\label{eq:R-tree}
   MBR(P)=\left[\min _{p \in P} p_{x}, \max _{p \in P} p_{x}\right] \times\left[\min _{p \in P} p_{y}, \max _{p \in P} p_{y}\right] 
\end{equation}
where  $p_{x}$  and  $p_{y}$ are the  $\mathrm{x}$  and  $\mathrm{y}$  coordinates of point  $p$ , and $\times$  denotes the Cartesian product.
The R-tree algorithm also uses two metrics to measure the quality of a node split: area and overlap. The area of a node is the area of its MBR, and the overlap of two nodes is the area of the intersection of their MBRs. The formula for computing the area of a node  N  is:
\vspace{-3mm}

\begin{equation}
\vspace{-1mm}
\small
\label{eq:R-tree2}
\operatorname{area}(N)=\left(N . x_{\max }-N . x_{\min }\right) \times\left(N \cdot y_{\max }-N \cdot y_{\min }\right)
\end{equation}
where  $N . x_{\min }$, $N . x_{\max }$, $N . y_{\min }$, and $N . y_{\max }$ are the coordinates of the MBR of node  $N$. 
The advantage of R-tree is that it can support spatial queries, such as range queries or nearest neighbor queries, on data points that represent geographical coordinates, rectangles, polygons, or other spatial objects. 
R-tree search performance depends on roughly the same factors as B-tree and also faces similar challenges as B-tree. 

\textbf{M-Tree}~\cite{ciaccia1997m}. It is a technique for finding the nearest neighbors of a given vector in a large collection of vectors. It works by building an M-tree, which is a tree data structure that partitions the data points into balls, i.e., hyperspheres that contain a subset of the points. Each node of the M-tree defines the smallest ball that contains all the points in its subtree. The algorithm then searches for the closest ball to the query point and then searches within the closest ball to find the closest point to the query point.

The M-tree algorithm uses the concept of covering radius to represent the spatial objects in the tree. The covering radius of a node is the maximum distance from the node's routing object to any of its child objects. The formula for computing the covering radius of a node  $N$  is:
\vspace{-3mm}

\begin{equation}
\vspace{-1mm}
\label{eq:M-tree1}
    r(N)=\max _{C \in N.child} d(N.object, C.object)
\end{equation}
where $N.object$ is the routing object of node  $N$, $N.child$ is the set of child nodes of node  $N$, $C.object$ is the routing object of child node  $C$, and $d$ is the distance function.

The M-tree algorithm also uses two metrics to measure the quality of a node split: area and overlap. The area of a node is the sum of the areas of its children's covering balls, and the overlap of two nodes is the sum of the areas of their children's overlapping balls. The formula for computing the area of a node $N$ is:
\vspace{-3mm}

\begin{equation}
\vspace{-1mm}
\label{eq:M-tree2}
\operatorname{area}(N)=\sum_{C \in \text { N.child }} \pi r(C)^{2}
\end{equation}
where  $\pi$  is the mathematical constant, and $r(C)$ is the covering radius of child node $C$.
The advantage of M-tree is that it can support dynamic operations, such as inserting or deleting data points, by updating the tree structure accordingly. M-tree search performance depends on roughly the same factors as B-tree, and also faces similar challenges as B-tree.

\subsubsection{Summary and Trade-offs of NNS Methods}
Exact nearest neighbor search guarantees returning the true nearest neighbor but faces performance challenges as data size and dimensionality grow. Brute-force search is simple and accurate but its linear complexity limits it to small datasets.

Tree-based methods (KD-Tree, Ball-Tree, R-Tree, M-Tree) reduce search space through spatial partitioning. KD-Tree works well in low dimensions but suffers from the curse of dimensionality. Ball-Tree handles high dimensions better at higher construction cost. R-Tree supports rectangle queries for spatial data but may require multiple search paths due to node overlap. M-Tree allows dynamic updates with higher maintenance overhead.
Exact methods are valuable for small datasets, low dimensions, or strict accuracy needs. In large-scale high-dimensional settings, they are often replaced by approximate methods.

\vspace{-3mm}
\subsection{Approximate Nearest Neighbor Search}
\subsubsection{Hash-Based Approach} 
\begin{figure}[]
	\centering
    \vspace{-3mm}
	\includegraphics[width=2in]{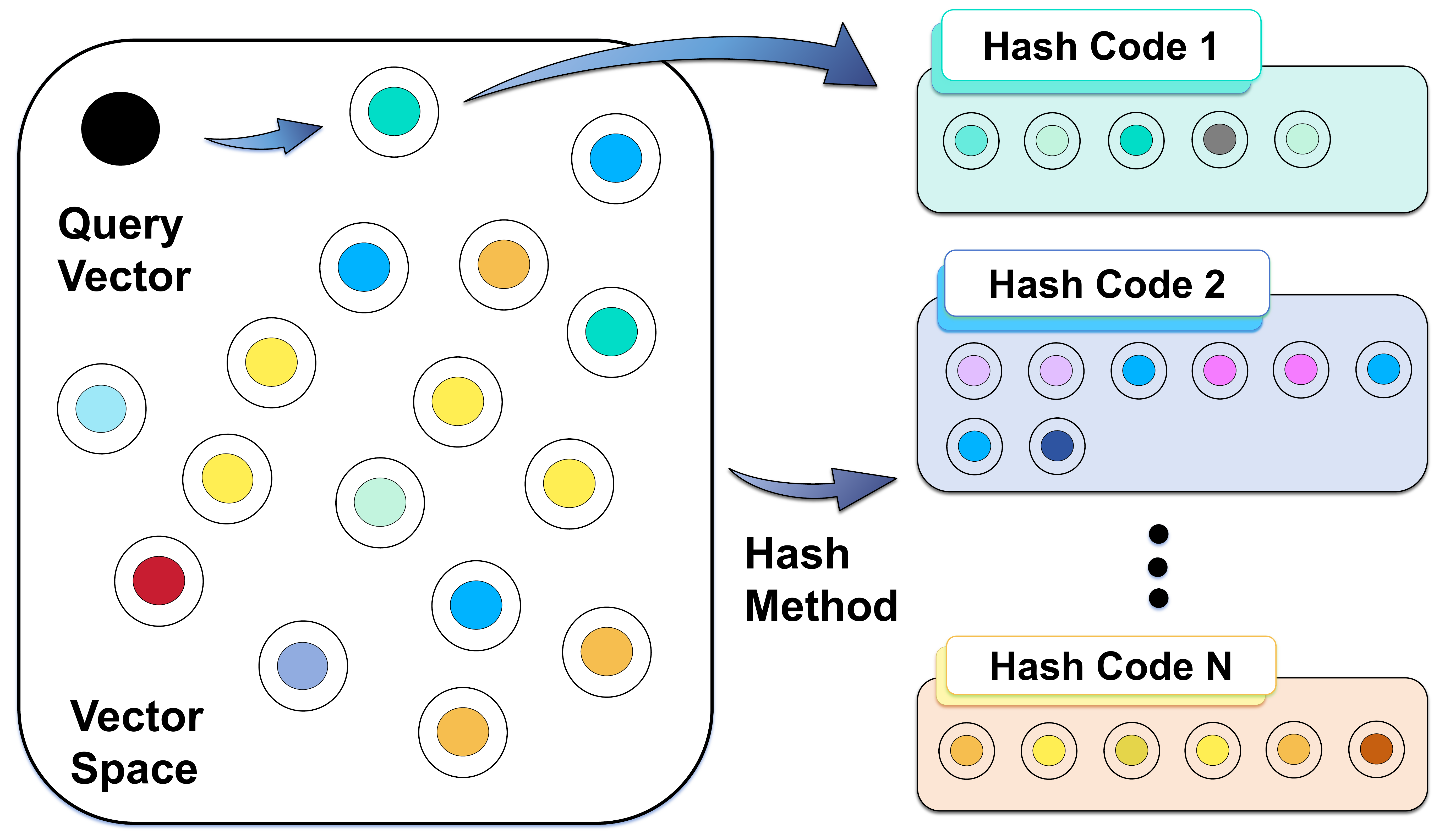}
    	\caption{The process of approximate nearest neighbor search based on hash approach}
	\label{fig:The process of approximate nearest neighbor search based on hash approach}
\end{figure}

The core idea of the hash-based approach is to reduce search complexity by mapping high-dimensional data to lower-dimensional hash codes with carefully designed hash functions while preserving similarity between data points. 
As shown in Figure~\ref{fig:The process of approximate nearest neighbor search based on hash approach}, each high-dimensional vector is transformed into a low-dimensional hash code. 
Similar points are mapped to the same or neighboring codes, so the search only needs to examine a small subset of codes, greatly improving efficiency. 
Based on this idea, four representative methods will be introduced: locality-sensitive hashing, spectral hashing, Spherical Hashing, and deep hashing. 

The idea is to reduce the memory footprint and the search time by comparing the binary codes instead of the original vectors~\cite{cai2019revisit}. 

\textbf{Local-Sensitive Hashing}~\cite{datar2004locality,jafari2021survey}. It is a technique for finding the approximate nearest neighbors of a given vector in a large collection of vectors. It works by using a hash function to transform the high-dimensional vectors into compact binary codes, and then using a hash table to store and retrieve the codes based on their similarity or distance. 
In LSH, hash functions are designed to preserve the locality of vectors. Unlike traditional hash functions, LSH increases the probability that similar items are mapped to the same code, thus increasing collisions among similar vectors.

A trace of algorithm description and implementation for locally sensitive hashing can be seen on the home page~\cite{andoni2023lsh}.

The LSH algorithm works by using a family of hash functions that use random projections or other techniques which are locality sensitive, meaning that similar vectors are more likely to have the same or similar codes than dissimilar vectors~\cite{dikkala2021manifold}, which satisfy the following property:

\begin{equation}
\label{eq:LSH1}
    \operatorname{Pr}[h(p)=h(q)]=f(d(p, q))
\end{equation}
where $h$ is a hash function, $p$  and  $q$  are two points, $d$ is a distance function, and $f$  is a similarity function. The similarity function $f$ is a monotonically decreasing function of the distance, such that the closer the points are, the higher the probability of collision.

There are different families of hash functions for different distance functions and similarity functions. For example, one of the most common families of hash functions for Euclidean distance and cosine similarity is:
\begin{equation}
\label{eq:LSH2}
    h(p)=\left\lfloor\frac{a \cdot p+b}{w}\right\rfloor
\end{equation}
where $a$ is a random vector, $b$ is a random scalar, and $w$ is a parameter that controls the size of the hash bucket. The similarity function for this family of hash functions is:

\begin{equation}
\label{eq:LSH3}
f(d(p, q))=1-\frac{d(p, q)}{\pi w},
\end{equation}
where $d(p, q)$ is the Euclidean distance between $p$ and  $q$. 
The advantage of LSH is that it can reduce the memory footprint and the search time by comparing the binary codes instead of the original vectors and also adapt to dynamic data sets by inserting or deleting codes from the hash table without affecting the existing codes~\cite{bob2022locality}. 
The performance of LSH depends on several factors, such as the dimensionality of the data, the number of hash functions, the number of bits per code, and the desired accuracy and recall. There are also some challenges and extensions of LSH, such as dealing with noisy and outlier data, choosing a good hash function family, or using multiple hash tables to increase recall. It is improved by \cite{andoni2008near,andoni_beyond_2013,andoni_optimal_2015}.

\textbf{Spectral Hashing}~\cite{weiss2008spectral}. 
It is a technique for finding the approximate nearest neighbors of a given vector in a large collection of vectors. It works by using spectral graph theory to generate hash functions that minimize the quantization error and maximize the variance of the binary codes. Spectral hashing can perform well when the data points lie on a low-dimensional manifold embedded in a high-dimensional space.

The spectral hashing algorithm works by solving an optimization problem that balances two objectives: (1) minimizing the variance of each binary function, which ensures that the data points are evenly distributed among the hypercubes, and (2) maximizing the mutual information between different binary functions, which ensures that the binary code is informative and discriminative. The optimization problem can be formulated as follows:
\vspace{-3mm}

\begin{equation}
\vspace{-1mm}
\label{eq:Spectral Hashing}
    \min _{y_{1}, \ldots, y_{n}} \sum_{i=1}^{n} Var\left(y_{i}\right)-\lambda I\left(y_{1}, \ldots, y_{n}\right)
\end{equation} 
where  $y_{i}$  is the  $i$-th binary function,  $Var\left(y_{i}\right)$  is its variance,  $I\left(y_{1}, \ldots, y_{n}\right)$  is the mutual information between all the binary functions, and  $\lambda$  is a trade-off parameter.

The advantage of spectral hashing is that it can perform well when the data points lie on a low-dimensional manifold embedded in a high-dimensional space. 
Spectral hashing search performance depends on roughly the same factors as local-sensitive hashing. There are also some challenges and extensions of spectral hashing, such as dealing with noisy and outlier data, choosing a good graph Laplacian for the data manifold, or using multiple hash functions to increase recall.

\textbf{Spherical Hashing}. 
Spherical hashing is a binary encoding technique based on hyperspheres, designed for efficient ANNS. 
Unlike traditional hyperplane-based methods, it partitions the data space using hyperspheres, which define tighter and more compact regions through their centers and radii. 
Each spherical hashing function, as described by Heo~\cite{heo2012spherical}, is characterized by $( p_k \in \mathbb{R}^D)$ and a distance threshold $t_k \in \mathbb{R}^{+}$, as detailed below: 

\vspace{-3mm}

\begin{equation}
\vspace{-1mm}
\label{eq:Spherical Hashing1}
 h_k(x)= \begin{cases}-1 & \text { when } d\left(p_k, x\right)>t_k \\ +1 & \text { when } d\left(p_k, x\right) \leq t_k\end{cases}
\end{equation}
where $d(\cdot,\cdot)$ is the Euclidean distance between two points in $D$-dimensional real space; however, alternative distance metrics, such as the Lp-norms, could also be employed in place of the Euclidean distance. 
The output of each spherical hashing function $h_{k}(x)$ determines if the point $x$ resides within the hypersphere that has $p_k$ as its center and $t_k$ as its radius. 
To improve similarity measurement, spherical hashing introduces the spherical Hamming distance, which accounts for the number of shared hyperspheres. 
The spherical Hamming distance is formulated as follows: 

\begin{equation}
\label{eq:Spherical Hashing2}
  d_{\text {shd }}\left(b_i, b_j\right)=\frac{\left|b_i \oplus b_j\right|}{\left|b_i \wedge b_j\right|}
\end{equation}
where $\left|b_i \oplus b_j\right|$ represents the number of different bits (where the XOR operation results in 1) between two binary codes, $\left|b_i \wedge b_j\right|$ represents the number of common bits (where the AND operation results in 1) between the two binary codes. 

Compared to hyperplane-based hashing functions, spherical hashing can map more spatially coherent data points into binary codes. 
Moreover, in high-dimensional spaces, hyperspheres are more powerful than hyperplanes in defining closed regions, allowing more potential nearest neighbors to be captured within the binary code region of a query point. 

\textbf{Deep Hashing}~\cite{liu2016deep,luo2022survey}. It is a technique for finding the approximate nearest neighbors of a given vector in a large collection of vectors. It works by using a deep neural network to learn hash functions that transform high-dimensional vectors into compact binary codes, and then using a hash table to store and retrieve the codes based on their similarity or distance~\cite{luo2023survey}. The hash functions are designed to preserve the semantic information of the vectors, which means that similar vectors are more likely to have the same or similar codes than dissimilar vectors~\cite{noauthor_remote_nodate}.

The deep hashing algorithm works by optimizing an objective function that balances two terms: (1) a reconstruction loss that measures the fidelity of the binary codes to the original data points and (2) a quantization loss that measures the discrepancy between the binary codes and their continuous relaxations. The objective function can be formulated as follows:

\vspace{-1mm}
\begin{equation}
\vspace{-1mm}
\label{eq:Deep Hashing}
   \min _{W, B} \sum_{i=1}^{N}\left\|x_{i}-W b_{i}\right\|_{2}^{2}+\lambda\left\|b_{i}-\operatorname{sgn}\left(b_{i}\right)\right\|_{2}^{2} 
\end{equation}
where  $x_{i}$  is the  $i$-th data point, $b_{i}$ is its continuous relaxation,  $\operatorname{sgn}\left(b_{i}\right)$ is its binary code, $W$ is a weight matrix that maps the binary codes to the data space, and $\lambda$ is a trade-off parameter.

The advantage of deep hashing is that it can leverage the representation learning ability of neural networks to generate more discriminative and robust codes for complex data, such as images, texts, or audio. 
The performance of deep hashing depends on several factors, such as the architecture of the neural network, the loss function used to train the network, and the number of bits per code. 

\subsubsection{Tree-Based Approach}
The main idea of the tree-based approach is to build hierarchical or recursively partitioned data structures, such as trees, to break high-dimensional datasets into smaller subsets. 
This method improves query efficiency by reducing the number of points that need to be searched. 
Along this line, three tree-based methods will be presented: approximate nearest neighbors oh yeah, best bin first, and k-means tree. 

The idea is to reduce the search space by following the branches of the tree that are most likely to contain the nearest neighbors of the query point. 

\textbf{Approximate Nearest Neighbors Oh Yeah}~\cite{bernhardsson2015annoy}.
It is a technique that can perform fast and accurate similarity searches and retrieval of high-dimensional vectors. It works by building a forest of binary trees, where each tree splits the vector space into two regions based on a random hyperplane. Each vector is then assigned to a leaf node in each tree based on which side of the hyperplane it falls on. To query a vector, Annoy traverses each tree from the root to the leaf node that contains the vector and collects all the vectors in the same leaf nodes as candidates. Then, it computes the exact distance or similarity between the query vector and each candidate and returns the top $k$ nearest neighbors. 
The formula for finding the median hyperplane between two points $p$ and $q$ is:

\begin{equation}
\label{eq:Approximate Nearest Neighbors Oh Yeah1}
   w \cdot x+b=0 
\end{equation}
where  $w=p-q$  is the normal vector of the hyperplane, $x$ is any point on the hyperplane, and $b=-\frac{1}{2}(w \cdot p+w \cdot q)$ is the bias term. 
The formula for assigning a point  x  to a leaf node in a tree is:
\begin{equation}
\label{eq:Approximate Nearest Neighbors Oh Yeah2}
\operatorname{sign}\left(w_{i} \cdot x+b_{i}\right)
\end{equation} 
where $w_{i}$ and $b_{i}$ are the normal vector and bias term of the  $i$-th split in the tree, and sign is a function that returns $1$ if the argument is positive, $-1$ if negative, and $0$ if zero. The point $x$ follows the left or right branch of the tree depending on the sign of this expression, until it reaches a leaf node.
The formula for searching for the nearest neighbor of a query point $q$ in the forest is:
\vspace{-3mm}

\begin{equation}
\vspace{-2mm}
\label{eq:Approximate Nearest Neighbors Oh Yeah3}
\min _{x \in C(q)} d(q, x)
\end{equation}
where $C(q)$ is the set of candidate points obtained by traversing each tree in the forest and retrieving all the points in the leaf node that $q$ belongs to, and $d$ is a distance function, such as Euclidean distance or cosine distance. The algorithm uses a priority queue to store the nodes to be visited, sorted by their distance to $q$. The algorithm also prunes branches that are unlikely to contain the nearest neighbor by using a bound on the distance between $q$ and any point in a node.

The advantage of Annoy is that it can use multiple random projection trees to index the data points, which can increase the recall and robustness of the search, also reduce the memory usage, and improve the speed of NNS by creating large read-only file-based data structures that are mapped into memory so that many processes can share the same data. 
The performance of Annoy depends on several factors, such as the dimensionality of the data, the number of trees built, the number of nearest candidates to search, and the distance approximation method used. 

\textbf{Best Bin First}~\cite{beis1997shape,liu2011improved}.
It is a technique for finding the approximate nearest neighbors of a given vector in a large collection of vectors. It works by building a kd-tree that partitions the data points into bins and then searching for the closest bin to the query point. The algorithm then searches within the closest bin to find the closest point to the query point. 
The best bin first algorithm still follows~\eqref{eq:kdtree-left} ~\eqref{eq:kdtree-right}.
The advantage of best bin first is that it can reduce the search time and improve the accuracy of NNS, by focusing on the most promising bins and avoiding unnecessary comparisons with distant points. 
The performance of the best bin first depends on several factors, such as the dimensionality of the data, the number of bins per node, the number of nearest candidates to search, and the distance approximation method used. 

\textbf{K-means Tree}~\cite{tavallali2021k}.
It is a technique for clustering high-dimensional data points into a hierarchical structure, where each node represents a cluster of points. It works by applying a k-means clustering algorithm to the data points at each level of the tree and then creating child nodes for each cluster. The process is repeated recursively until a desired depth or size of the tree is reached. 
The formula for assigning a point $x$ to a cluster using the k-means algorithm is:
\begin{equation}
\label{eq:K-means tree1}
    \mathop{\arg\min}_{i=1, \ldots, k}\left\|x-c_{i}\right\|_{2}^{2}
\end{equation}
where argmin is a function that returns the argument that minimizes the expression and  $\|\cdot\|_{2}$  denotes the Euclidean norm. 
The formula for assigning a point $x$ to a leaf node in a k-means tree is:
\begin{equation}
\label{eq:K-means tree2}
\mathop{\arg\min}_{N \in L(x)} \| x-N.center\|_{2}^{2}
\end{equation}
where $L(x)$ is the set of leaf nodes that  $x$  belongs to, and  $N.center$ is the cluster center of node $N$. The point $x$ belongs to a leaf node if it belongs to all its ancestor nodes in the tree. 
The formula for searching for the nearest neighbor of a query point $q$ in the k-means tree is:
\begin{equation}
\label{eq:K-means tree3}
\min _{x \in C(q)}\|q-x\|_{2}^{2}
\end{equation}
where $C(q)$ is the set of candidate points obtained by traversing each branch of the tree and retrieving all the points in the leaf nodes that $q$ belongs to. The algorithm uses a priority queue to store the nodes to be visited, sorted by their distance to $q$. The algorithm also prunes branches that are unlikely to contain the nearest neighbor by using a bound on the distance between $q$ and any point in a node.

The advantage of K-means tree is that it can perform fast and accurate similarity searches and retrieval of data points based on their cluster membership by following the branches of the tree that are most likely to contain the nearest neighbors of the query point. K-means tree can also support dynamic operations, such as inserting and deleting points, by updating the tree structure accordingly. 
The performance of K-means tree depends on several factors, such as the dimensionality of the data, the number of clusters per node, and the distance metric used. 

\subsubsection{Graph-Based Approach}
The core concept of the graph-based approach is the small-world network. 
A small-world network is a complex network where most nodes are not directly connected, but almost any node can be reached from another in just a few steps. 
This means the average path length between any two nodes is much shorter than the total number of nodes. 
The "six degrees of separation" theory in sociology~\cite{guare2016six} is a concrete manifestation of a small-world network. 

\begin{figure}[]
    \centering
    \vspace{-3mm}
    \includegraphics[width=3in]{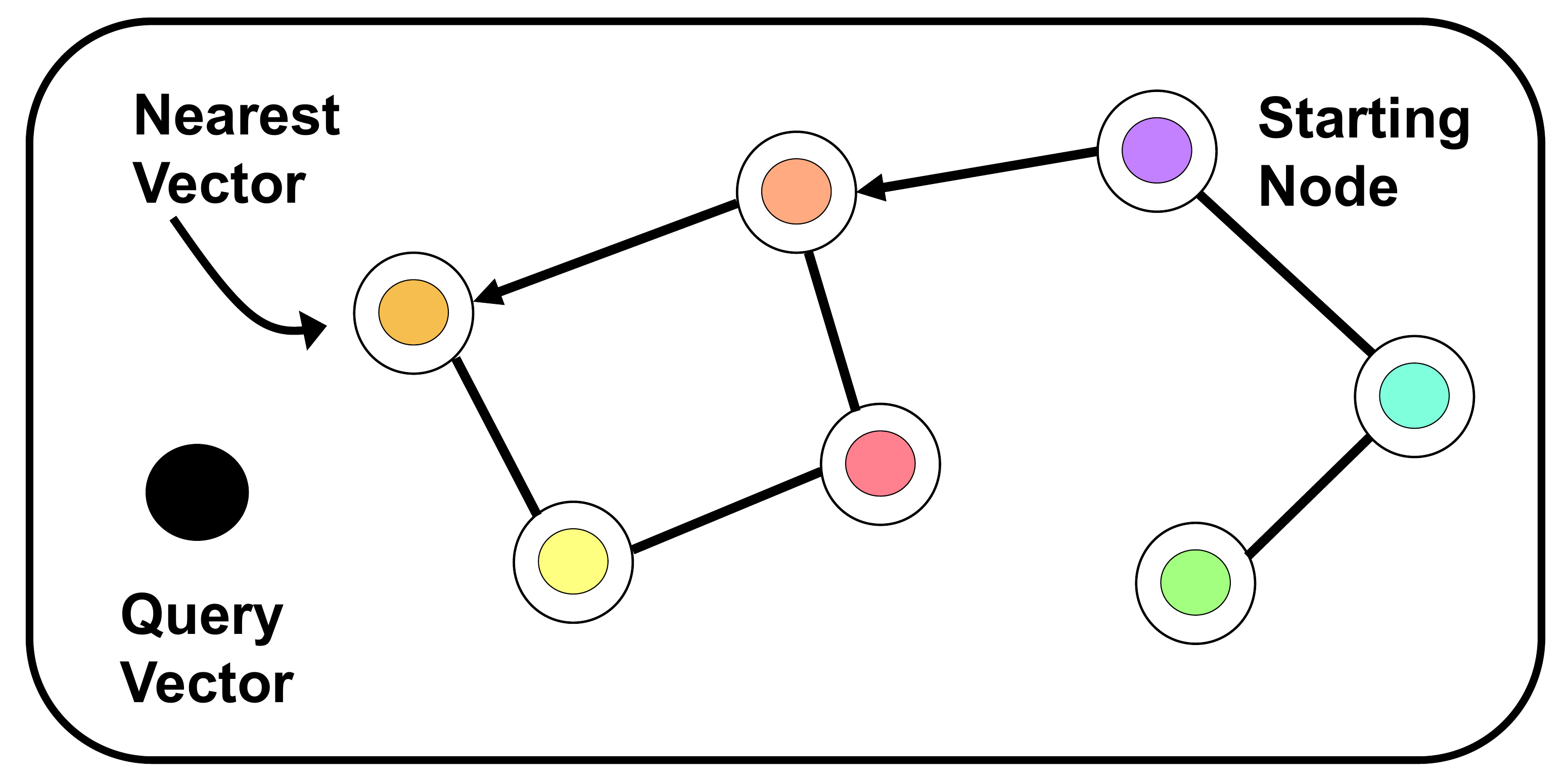}
    \vspace{-2mm}
	\caption{The process of performing nearest neighbor search on a small-world network.
    }
    \vspace{-3mm}
    
	\label{fig:The process of performing nearest neighbor search on a small-world network}
\end{figure}

Given a query vector $q$ and the task of finding the $k$ closest vectors from a set $O$, the graph-based approach uses a graph $G(V, E)$ to represent these objects, where each object $o_i$ corresponds to a node $v_i$. 
The small-world network graph is constructed by sequentially adding all nodes. 
As shown in Figure~\ref{fig:The process of performing nearest neighbor search on a small-world network}, when a new node is added, a list of neighboring nodes is generated using a greedy algorithm, and bidirectional connections are established between the new node and all nodes in the list. 
Once the graph is constructed, searching for $q$ is similar to the process of adding a new node. 
The search starts from a randomly selected node (with different small-world variants possibly using different selection strategies). 
From the list of neighbors of the current node, the node most similar to $q$ is identified. 
If such a node is found, it becomes the next node, and the search process is repeated. 
If a node has a higher similarity to $q$ than all of its neighbors, the search stops, and that node is considered the one most similar to $q$. 
Two types of graph-based methods are introduced: navigable small world (NSW) and hierarchical navigable small world (HNSW). 

\textbf{Navigable Small World} 
It is a technique that uses a graph structure to store and retrieve high-dimensional vectors based on their similarity or distance~\cite{ponomarenko2011approximate}. The NSW algorithm builds a graph by connecting each vector to its nearest neighbors, as well as some random long-range links that span different regions of the vector space. The idea is that these long-range links create shortcuts that allow for faster and more efficient traversal of the graph, similar to how social networks have small world properties~\cite{malkov2014approximate}.

The NSW algorithm works by using a greedy heuristic to add edges to the graph\cite{malkov2012scalable}. The algorithm starts with an empty graph and adds one point at a time. For each point, the algorithm finds its nearest neighbor in the graph using a random walk and connects it with an edge. Then, the algorithm adds more edges by connecting the point to other points that are closer than its current neighbors. The algorithm repeats this process until all points are added to the graph. 
The formula for finding the nearest neighbor of a point  p  in the graph using a random walk is:

\begin{equation}
    \mathop{\arg\min}_{q \in N(p)} d(p, q)
    \label{eq:nsw1}
\end{equation}
where  N(p)  is the set of neighbors of  p  in the graph, and  d  is a distance function, such as Euclidean distance or cosine distance. The algorithm starts from a random point in the graph and moves to its nearest neighbor until it cannot find a closer point.
The formula for adding more edges to the graph using a greedy heuristic is:

\begin{equation}
\begin{aligned}
&\forall q \in N(p), \forall r \in N(q) \text{, if } d(p, r) < d(p, q) \text{,} \\
&\text{then add edge } (p, r)
\end{aligned}
\label{eq:nsw2}
\end{equation}
where $N(p)$ and $N(q)$ are the sets of neighbors of $p$ and $q$ in the graph, respectively, and $d$ is a distance function. The algorithm connects $p$ to any point that is closer than its current neighbors.

The advantage of the NSW algorithm is that it can handle arbitrary distance metrics, it can adapt to dynamic data sets, and it can achieve high accuracy and recall with low memory consumption. The NSW algorithm also uses a greedy routing strategy, which means that it always moves to the node that is closest to the query vector until it reaches a local minimum or a predefined number of hops. 
The performance of the NSW algorithm depends on several factors, such as the dimensionality of the vectors, the number of neighbors per node, the number of long-range links per node, and the number of hops per query. 

\textbf{Hierachical Navigable Small World}~\cite{malkov2018efficient}.
It is a state-of-the-art technique for finding the approximate nearest neighbors of a given vector in a large collection of vectors. It works by building a graph structure that connects the vectors based on their similarity or distance and then using a greedy search strategy to traverse the graph and find the most similar vectors.
The HNSW algorithm still follows~\eqref{eq:nsw1} and~\eqref{eq:nsw2}.
The HNSW algorithm also builds a hierarchical structure of the graph by assigning each point to different layers with different probabilities. The higher layers contain fewer points and edges, while the lower layers contain more points and edges. When a search query comes in, the HNSW algorithm finds the closest matching data points in the highest layer. It then proceeds layer by layer, moving downwards and finding the nearest data points in each subsequent layer based on those from the layer above. These points are considered the nearest neighbors. The algorithm continues this process in the lower layers, updating the list of nearest neighbors at each step. Once it reaches the bottom layer, the HNSW algorithm returns the data points that are closest to the search query. The algorithm uses a parameter M to control the maximum number of neighbors for each point in each layer. 
The formula for assigning a point $p$ to a layer $l$ using a random probability is:

\begin{equation}
\label{eq:HNSW1}
    \operatorname{Pr}[p \in l]=\left\{\begin{array}{cc}
1 & \text { if } l=0 \\
\frac{1}{M} & \text { if } l>0
\end{array}\right.
\end{equation}
where $M$ is the parameter that controls the maximum number of neighbors for each point in each layer. The algorithm assigns  $p$ to layer $l$ with probability $\operatorname{Pr}[p \in l]$, and stops when it fails to assign $p$ to any higher layer. 
The formula for searching for the nearest neighbor of a query point  q  in the hierarchical graph is:
\begin{equation}
\label{eq:HNSW2}
\min _{p \in C(q)} d(q, p)
\end{equation}
where $C(q)$ is the set of candidate points obtained by traversing each layer of the graph from top to bottom and retrieving all the points that are closer than the current best distance. The algorithm uses a priority queue to store the nodes to be visited, sorted by their distance to $q$. The algorithm also prunes branches that are unlikely to contain the nearest neighbor by using a bound on the distance between $q$ and any point in a node.

The advantage of HNSW is that it can achieve better performance than other methods of ANNS, such as tree-based or hash-based techniques. For example, it can handle arbitrary distance metrics, it can adapt to dynamic data sets, and it can achieve high accuracy and recall with low memory consumption. The performance of HNSW depends on several factors, such as the dimensionality of the vectors, the number of layers, the number of neighbors per node, and the number of hops per query.

Beyond the greedy traversal inherent to graph-based methods, further acceleration can be achieved through geometric pruning techniques that leverage metric space properties. Triangle inequalities, for example, can safely eliminate nodes unlikely to be closer than the current best candidate during graph search, reducing distance computations. This principle extends beyond graphs to other index structures. Recent work demonstrates this by applying fine-grained triangle inequality pruning within cluster-based indexes, achieving over 99.4\% pruning ratio\cite{xu2025tribase}. Such geometric pruning strategies are orthogonal to the underlying index structure—whether graph, tree, or cluster-based—and can be integrated for compounded performance gains.

\subsubsection{Quantization-Based Approach}
\begin{figure}[thbp]
	\centering
    \vspace{-3mm}
	\includegraphics[width=3in]{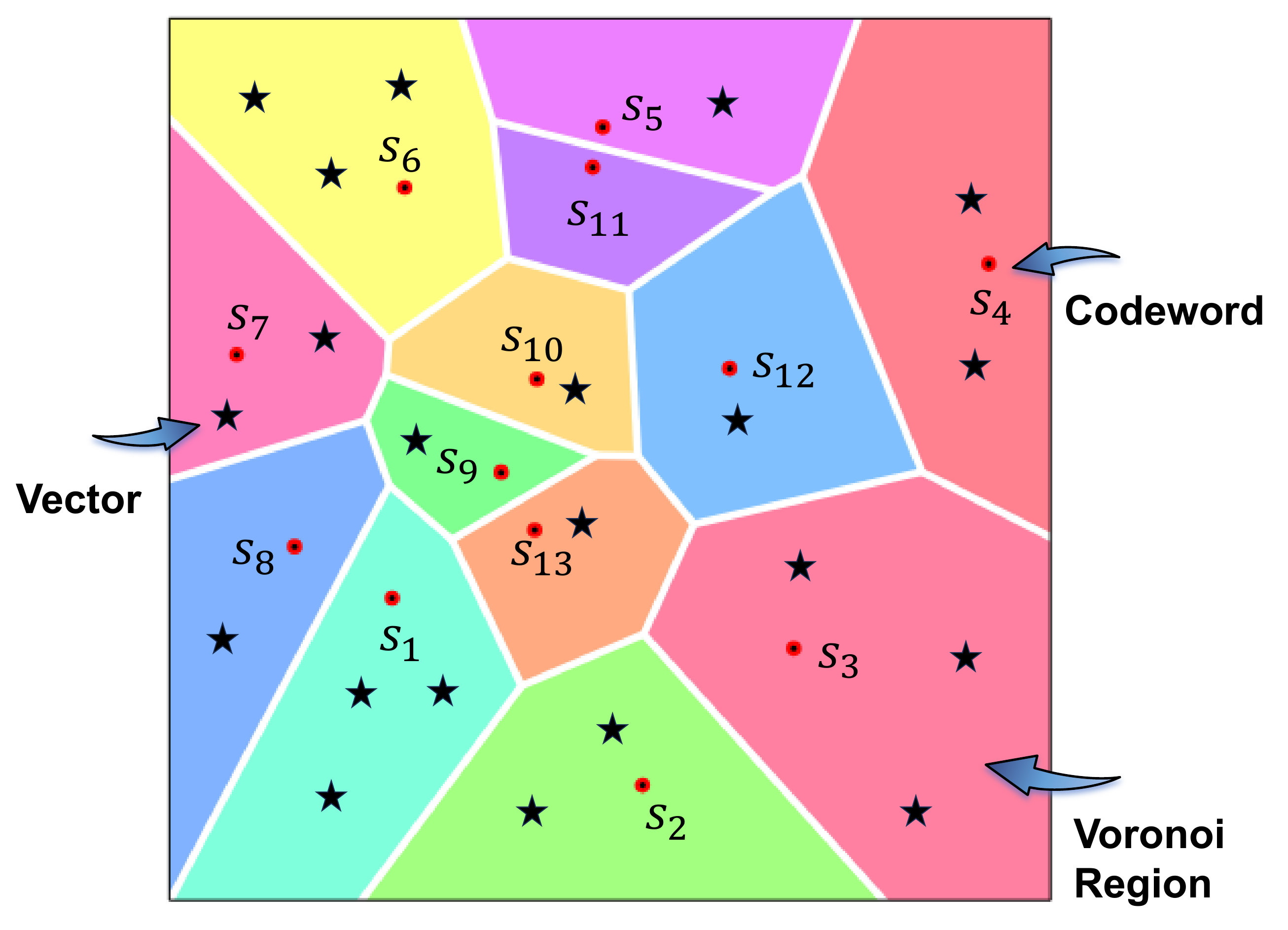}
    \vspace{-2mm}
    
	\caption{Codewords and Voronoi Regions in a Two-Dimensional Space}
    \vspace{-3mm}
	\label{fig:Codewords and Voronoi Regions in a Two-Dimensional Space}
\end{figure}

The core idea of quantization is to map points in a high-dimensional space to a low-precision representation in a finite set. 
This reduces the number of bits required for storage, lowering storage demands. 
By using these quantized representations during queries, the distance between the original points can be quickly estimated. 
Specifically, a vector quantizer maps k-dimensional vectors from the vector space $R^{k}$ to a finite set of vectors $S=\{s_{i}:i=1,\dots,n\}$. Each vector $s_{i}$ is called a code vector or codeword, or centroids. The collection of all codewords is referred to as a codebook. Associated with each codeword, $s_{i}$, is a nearest neighbor region called Voronoi region, and it is defined by: $V_i=\left\{x \in R^k:\left\|x-y_i\right\| \leq\left\|x-y_j\right\|, \text { forall } j \neq i\right\}$. The set of Voronoi region partitions the entire space $R^{k}$ such that: $\bigcup_{i=1}^N V_i=R^k$,
$\bigcap_{i=1}^N V_i=\phi \text { for all } i \neq j$. 
Figure~\ref{fig:Codewords and Voronoi Regions in a Two-Dimensional Space} shows the codewords in a two-dimensional space, where input vectors are marked with blue stars, codewords are marked with orange circles, and the Voronoi regions are separated with boundary lines.

\begin{figure}[thbp]
	\centering
	\includegraphics[width=3in]{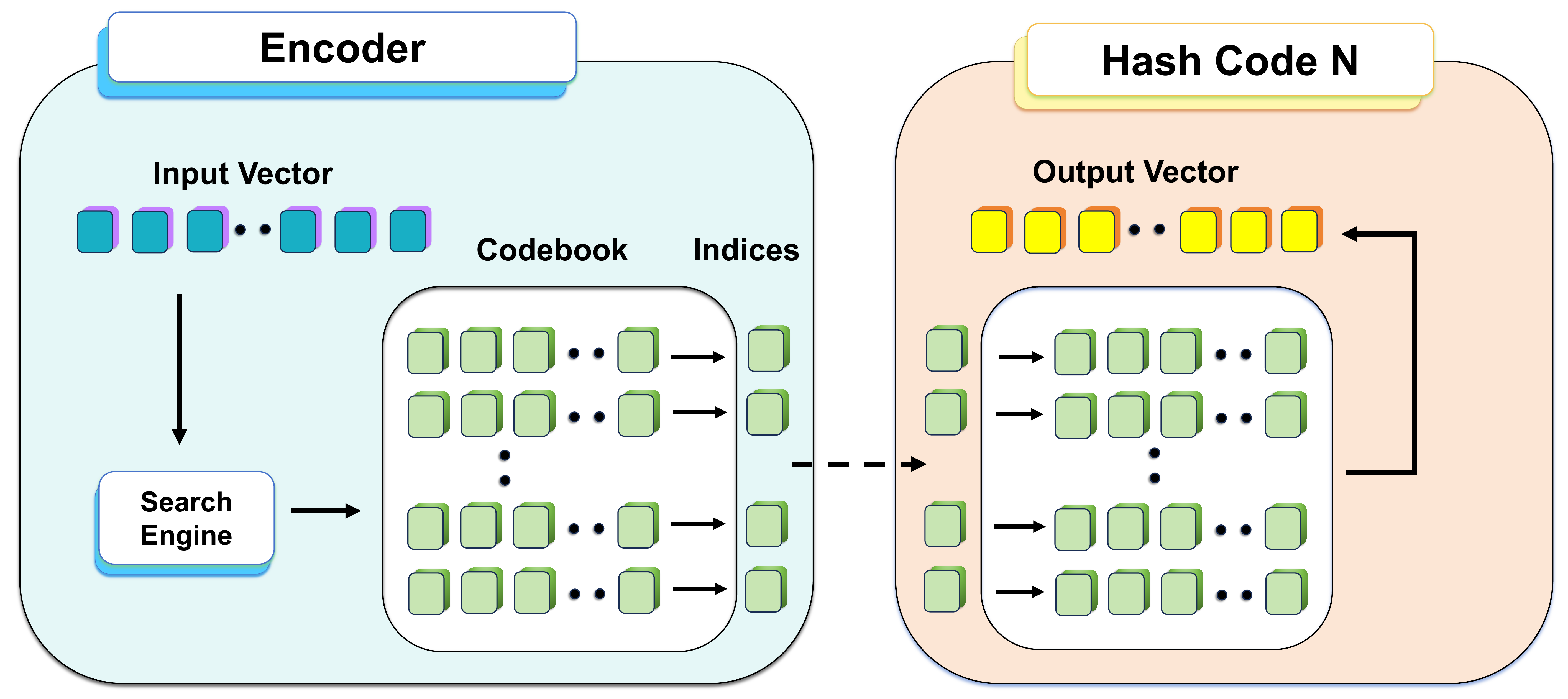}
	\caption{Vector Quantization Process with Encoder and Decoder Operations}
    
	\label{Vector Quantization Process with Encoder and Decoder Operations}
    \vspace{-3mm}
\end{figure}

A vector quantizer consists of two primary components: an encoder and a decoder, as shown in Figure~\ref{Vector Quantization Process with Encoder and Decoder Operations}. 
The encoder takes an input vector and outputs the index of the codeword that minimizes the distortion. 
This minimal distortion is found by calculating the distance between the input vector and each codeword in the codebook, typically using metrics like Euclidean or Hamming distance. Once the codeword with the smallest distance is identified, its index is transmitted to the decoder. 
At the receiver, the decoder then maps this index back to the corresponding codeword.
Generating an effective codebook involves selecting codewords that best represent a given set of input vectors, along with determining the appropriate number of codewords. Designing an optimal codebook is an NP-hard problem, implying that finding the absolute best set of codewords through exhaustive search becomes impractically complex as the number of codewords increases. As a result, heuristic methods, such as the Linde-Buzo-Gray (LBG) algorithm, which is conceptually similar to the K-means clustering algorithm, are commonly employed.
To create a codebook using the LBG algorithm, one first specifies the number of codewords, $N$, which defines the size of the codebook. 
Initially, $N$ codewords are selected at random, often from the set of input vectors themselves. Each input vector is then associated with the nearest codeword based on the Euclidean distance. After all vectors have been assigned to their respective clusters, a new set of codewords is generated by computing the average of the vectors within each cluster. 
This process involves summing the components of the vectors in each cluster and dividing by the total number of vectors in that cluster. This iterative procedure refines the codebook until a satisfactory level of representation is achieved.
The formula for calculating the average of the components within each cluster is:
$
y_i=\frac{1}{m} \sum_{j=1}^m x_{i j}
$,
where $i$ is the component of each vector and $m$ is the number of vectors in the cluster. There are also many other methods for designing the codebook, methods such as Generative Pre-trained Transformer Vector quantization (GPTVQ)~\cite{van2024gptvq}, Vector PostTraining Quantization (VPTQ)~\cite{liu2024vptq}, deep network architecture for vector quantization (DeepVQ)~\cite{le2018deepvq}, etc. 
Based on the core idea described above, six quantization-based methods will be presented here, namely inverted file index (IVF), product quantization (PQ)~\cite{jegou2010product,matsui2018survey}, optimized product quantization (OPQ)~\cite{ge2013optimized,li2020optimized}, online product quantization\cite{noauthor_online_nodate}, scalable nearest neighbor (ScaNN)~\cite{guo2020accelerating}, and inverted file product quantization (IVF\_PQ)~\cite{guo2022manu,wang2021milvus}. 
Product quantization can reduce the memory footprint and search time of ANN search by comparing codes instead of the original vectors ~\cite{8653810}.  More recently, RaBitQ \cite{gao2024rabitq}, a randomized quantization method that provides strong theoretical guarantees on distance estimation error while maintaining practical efficiency. These approaches are fundamental to modern high-dimensional similarity search.

\textbf{Inverted File Index}. Inverted File Index is a technique designed to enhance search efficiency by narrowing the search area through the use of neighbor partitions or clusters~\cite{liu2022new}. 
It uses clustering (e.g., K-means) to partition high-dimensional vectors into multiple regions (Voronoi Cells) and records the vectors within each region through an inverted index. During a query, the search is restricted to a few regions closest to the query vector, significantly reducing the search space and improving retrieval efficiency. 
IVF is often combined with other techniques, such as Product Quantization (PQ), to further optimize storage and computation, making it widely used in image retrieval, recommendation systems, and VDBs. 
Its main advantages are fast search speed and high efficiency, though its performance in high-dimensional spaces may be limited by clustering quality and the complexity of dynamic updates.

\textbf{Product Quantization}.
It is a technique for compressing high-dimensional vectors into smaller and more efficient representations~\cite{jegou2010product,matsui2018survey}. It works by dividing a vector into several sub-vectors, and then applying a clustering algorithm (such as k-means) to each sub-vector to assign it to one of a finite number of possible values (called centroids). The result is a compact code that consists of the indices of the centroids for each sub-vector.

The PQ algorithm works by using a vector quantization technique to map each subvector to its nearest centroid in a predefined codebook. The algorithm first splits each vector into $m$ equal-sized subvectors, where $m$ is a parameter that controls the length of the code. Then, for each subvector, the algorithm learns $k$ centroids using the k-means algorithm, where $k$ is a parameter that controls the size of the codebook. Finally, the algorithm assigns each subvector to its nearest centroid and concatenates the centroid indices to form the code. 
The formula for splitting a vector $x$ into $m$ subvectors is:

\begin{equation}
\label{eq:PQ1}
    x=\left(x_{1}, x_{2}, \ldots, x_{m}\right)
\end{equation}
where $x_{i}$ is the  $i$-th subvector of $x$, and has dimension  $d/m$, where $d$ is the dimension of $x$. 
The formula for finding the centroids of a set of subvectors $P$ using the k-means algorithm is:

\begin{equation}
\label{eq:PQ2}
    c_{i}=\frac{1}{\left|S_{i}\right|} \sum_{x \in S_{i}} x
\end{equation}
where $c_{i}$ is the  $i$-th centroid, $S_{i}$ is the set of subvectors assigned to the  i-th cluster, and $|\cdot|$ denotes the cardinality of a set. 
The formula for assigning a subvector $x$ to a centroid using the k-means algorithm is:
\begin{equation}
\label{eq:PQ3}
\operatorname{argmin}_{i=1, \ldots, k}\left\|x-c_{i}\right\|_{2}^{2}
\end{equation}
where argmin is a function that returns the argument that minimizes the expression and  $\|\cdot\|_{2}$  denotes the Euclidean norm. 
The formula for encoding a vector $x$ using PQ is:
\begin{equation}
\label{eq:PQ4}
c(x)=\left(q_{1}\left(x_{1}\right), q_{2}\left(x_{2}\right), \ldots, q_{m}\left(x_{m}\right)\right)
\end{equation}
where $x_{i}$  is the $i$-th subvector of  $x$, and $q_{i}$ is the quantization function for the  $i$-th subvector, which returns the index of the nearest centroid in the codebook.

The advantage of product quantization is that it is simple and easy to implement, as it only requires a standard clustering algorithm and a simple distance approximation method. 
The performance of product quantization depends on several factors, such as the dimensionality of the data, the number of sub-vectors, the number of centroids per sub-vector, and the distance approximation method used.

\textbf{Optimized Product Quantization} ~\cite{ge2013optimized}.
It is a variation of product quantization (PQ), which is a technique for compressing high-dimensional vectors into smaller and more efficient representations. OPQ works by optimizing the space decomposition and the codebooks to minimize quantization distortions. OPQ can improve the performance of PQ by reducing the loss of information and increasing the discriminability of the codes ~\cite{li2020optimized}. 
The advantage of OPQ is that it can achieve higher accuracy and recall than PQ, as it can better preserve the similarity or distance between the original vectors. 
The formula for applying a random rotation to the data is:
\begin{equation}
\label{eq:OPQ1}
    x^{\prime}=R x
\end{equation}
where $x$ is the original vector, $x^{\prime}$ is the rotated vector, and $R$ is a random orthogonal matrix. The formula for finding the rotation matrix for a subvector using an optimization technique is:
\begin{equation}
\label{eq:OPQ2}
\min _{R_{i}} \sum_{x \in P_{i}}\left\|x-R_{i} c_{i}\left(R_{i} x\right)\right\|_{2}^{2}
\end{equation}
where $P_{i}$ is the set of subvectors assigned to the  $i$-th cluster, $R_{i}$ is the rotation matrix for the  $i$-th cluster, and $c_{i}$ is the quantization function for the $i$-th cluster, which returns the nearest centroid in the codebook. 
The formula for encoding a vector $x$ using  OPQ  is:
\begin{equation}
\label{eq:OPQ3}
c(x)=\left(q_{1}\left(R_{1} x_{1}\right), q_{2}\left(R_{2} x_{2}\right), \ldots, q_{m}\left(R_{m} x_{m}\right)\right)
\end{equation}
where  $x_{i}$  is the  $i$-th subvector of  $x$, $R_{i}$ is the rotation matrix for the $i$ -th subvector, and $q_{i}$ is the quantization function for the $i$-th subvector, which returns the index of the nearest centroid in the codebook.

The performance of OPQ depends on several factors, such as the dimensionality of the data, the number of sub-vectors, the number of centroids per sub-vector, and the distance approximation method used. 

\textbf{Online Product Quantization} ~\cite{xu2018online}.
It is a variation of product quantization (PQ), which is a technique for compressing high-dimensional vectors into smaller and more efficient representations. Online product quantization (O-PQ) works by adapting to dynamic data sets by updating the quantization codebook and the codes online. O-PQ can handle data streams and incremental data sets without requiring offline retraining or reindexing.
The formula for splitting a vector $x$ into $m$ subvectors is:

\begin{equation}
\label{eq:O-PQ1}
    x=\left(x_{1}, x_{2}, \ldots, x_{m}\right)
\end{equation}
where $x_{i}$ is the  $i$-th subvector of $x$, and has dimension  $d/m$, where $d$ is the dimension of $x$.
The formula for initializing the centroids of a set of subvectors $P$ using the k-means++ algorithm is:
\begin{equation}
\label{eq:O-PQ2}
c_{i}=\text { randomly choose a point from } P
\end{equation}
where $c_{i}$ is the  $i$-th centroid, with probability proportional to $D(x)^{2}$, $D(x)$ is the distance between point $x$ and its closest centroid among $\left\{c_{1}, \ldots, c_{i-1}\right\}$.
The formula for assigning a subvector $x$ to a centroid using $\mathrm{PQ}$ is:
\begin{equation}
\label{eq:O-PQ3}
\operatorname{argmin}_{i=1, \ldots, k}\left\|x-c_{i}\right\|_{2}^{2}
 \end{equation}
where $\arg\min$ is a function that returns the argument that minimizes the expression and $\|\cdot\|_{2}$  denotes the Euclidean norm. 
The formula for encoding a vector $x$ using $\mathrm{PQ}$ is:
\begin{equation}
\label{eq:O-PQ4}
c(x)=\left(q_{1}\left(x_{1}\right), q_{2}\left(x_{2}\right), \ldots, q_{m}\left(x_{m}\right)\right)
 \end{equation}
where $x_{i}$ is the  $i$-th subvector of $x$, and $q_{i}$ is the quantization function for the $i$-th subvector, which returns the index of the nearest centroid in the codebook.

The O-PQ algorithm also updates the codebooks and codes for each subvector using an online learning technique. The algorithm uses two parameters:  $\alpha$, which controls the learning rate, and  $\beta$, which controls the forgetting rate. The algorithm updates the codebooks and codes as follows: 
For each new point  $x$, assign it to its nearest centroid in each subvector using PQ. 
For each subvector $x_{i}$, update its centroid $c_{q_{i}\left(x_{i}\right)}$ as:
\begin{equation}
\label{eq:O-PQ5}
    c_{q_{i}\left(x_{i}\right)}=(1-\alpha) c_{q_{i}\left(x_{i}\right)}+\alpha x_{i}
\end{equation} 
For each subvector $x_{i}$, update its code $q_{i}\left(x_{i}\right)$ as:
\begin{equation}
\label{eq:O-PQ6}
q_{i}\left(x_{i}\right)=\arg\min_{j=1, \ldots, k}\left\|(1-\beta) x_{i}+\beta x_{i}-(1-\beta) c_{j}+\beta c_{j}\right\|_{2}^{2}
\end{equation}
where $x_{i}$ and  $c_{j}$ are the mean vectors of all points and centroids in subvector $i$, respectively.

The advantage of O-PQ is that it can deal with changing data distributions and new data points, as it can update the codebooks and the codes in real time. 
O-PQ search performance depends on roughly the same factors as OPQ and also faces similar challenges as OPQ. 

\textbf{Scalable Nearest Neighbor}. 
It is a technique for efficient vector similarity search at scale~\cite{guo2020accelerating,sun2023soar}. 
ScaNN optimizes Maximum Inner Product Search (MIPS) through search space pruning and quantization. Traditional MIPS schemes aim to minimize the average distance between each vector $x$ and its centroids $\tilde{x}$, that is, to minimize quantization distortions. 
The formula for typically measuring the quantization distortion is:
\begin{equation}
\label{eq:Scalable Nearest Neighbor1}
D=\frac{1}{N} \sum_{i=1}^N\left\|x_i-\tilde{x}_i\right\|^2_{2}
\end{equation}
where $N$ is the total number of vectors, $x_i$ is the original vector, 
$\tilde{x}$ is quantized centroid, and 
 $\|\cdot\|_{2}$ denotes Euclidean
norm. 
 
While the ScaNN algorithm argues that optimizing the average distance is not equivalent to optimizing the accuracy of nearest-neighbor searches. 
The hypothesis it puts forward is that the objective of maximizing the inner product between two points is not entirely consistent with the objective of minimizing the average distance between two points. 

ScaNN takes into account the distribution characteristics of the data in different directions; ellipsoidal or other shaped regions are used instead of spherical regions around the centroids to better fit the local structure of the data. 
Building on this perspective, the anisotropic loss function can further enhance the adaptability of vector quantization to data anisotropy. 
By explicitly separating quantization errors into parallel and orthogonal components, the anisotropic loss function assigns distinct scaling parameters $h_i^{\|}$ and $h_i^{\perp}$ to these components, respectively. 
This allows for more fine-grained control over the quantization process, ensuring that the errors are distributed in alignment with the data's geometric characteristics. 

The anisotropic vector quantization algorithm shares similarities with the Lloyd algorithm, iteratively refining the codebook and data partitions. The key distinction lies in the update rule for the codebook centroids:

\begin{equation}
\label{eq:Scalable Nearest Neighbor2}
    \mathbf{c}_j=\frac{\sum_{i \in X_j} h_i^{\|} \cdot \mathbf{x}_i^{\|}+h_i^{\perp} \cdot \mathbf{x}_i^{\perp}}{\sum_{i \in X_j}\left(h_i^{\|}+h_i^{\perp}\right)}
\end{equation}
where $X_j$  is the set of data points assigned to the codeword $\mathbf{c}_j$. This update formula takes into account the directional scaling, ensuring that the resulting codewords are optimally positioned in accordance with the anisotropic properties of the data. 

By integrating this anisotropic loss framework, the quantization process moves beyond spherical symmetry and better accommodates ellipsoidal or irregularly shaped distributions in the data. 
The performance of ScaNN depends on several factors, such as the anisotropy of the data distribution, the choice of quantization methods like vector or product quantization, the size and quality of the codebooks, and the efficiency of the partitioning and scoring processes. 

\textbf{Inverted File Product quantization}. 
It is a widely used technique for approximate nearest neighbor (ANN) search in high-dimensional vector spaces~\cite{guo2022manu,wang2021milvus}. This algorithm is a combination of the Inverted File Indexing (IVF) and Product Quantization (PQ) algorithms. IVF\_PQ first uses the IVF algorithm to divide or partition the data into clusters and uses the parameter $nprobe$ to control the number of clusters. The higher the $nprobe$, the better the search results, but it also increases the time required. 
It then identifies the top-N clusters closest to the query vector and performs the search within these N clusters using the Product Quantization (PQ) algorithm. 

The IVF\_PQ algorithm naturally results in two different approaches when using the PQ algorithm: the first involves performing K-means clustering with the IVF algorithm, followed by applying a local PQ algorithm for dimensionality reduction within each cluster; the second also starts with the IVF algorithm to divide all data points into several clusters but applies a globally unified PQ algorithm for dimensionality reduction within each cluster. 

\textbf{Randomized Bit Quantization}\cite{gao2024rabitq}. It is a recently proposed quantization method that addresses a key limitation of existing approaches: the lack of theoretical guarantees on distance estimation error, which can lead to catastrophic failures on some real-world datasets. It compresses D-dimensional vectors into D-bit strings while providing a sharp theoretical bound on the approximation error, ensuring that distance estimates are provably close to true distances. Beyond its theoretical guarantees, RaBitQ achieves excellent empirical performance through efficient implementations leveraging bitwise operations or SIMD-based acceleration. Extensive experiments show that RaBitQ outperforms PQ and its variants in accuracy-efficiency trade-off by a clear margin, with empirical behavior closely aligning with theoretical analysis. This combination of rigorous guarantees and practical efficiency makes RaBitQ particularly valuable for applications requiring both reliability and high performance.

\subsubsection{Summary and Trade-offs of ANNS Methods}
Four mainstream approximate nearest neighbor search methods exist: hashing-based, tree-based, graph-based, and quantization-based approaches. Each offers different trade-offs in accuracy, speed, memory, and dynamic update support.

Hashing-based methods (e.g., LSH) suit static high-dimensional sparse data but often have limited recall. Tree-based methods (e.g., KD-Tree) perform well in low dimensions but degrade in high dimensions due to the curse of dimensionality. Graph-based methods (e.g., HNSW) balance accuracy and speed effectively, making them popular in production, though they require high memory and careful maintenance. Quantization-based methods (e.g., PQ) compress vectors to reduce storage, ideal for memory-constrained settings, but may sacrifice accuracy.

Choosing the right method requires balancing data dimensionality, query patterns, hardware resources, and update frequency.

\section{Vector Database Comparison}

In the realm of VDBs, a variety of storage and search technologies has given rise to a diverse range of commercial and open-source solutions. In this section, to help users better understand the performance of different VDBs, we have conducted a comprehensive comparison of several popular options, including PgVector\footnote{http://github.com/pgvector.}, QdrantCloud\footnote{https://www.qdrant.tech.}, WeaviateCloud\footnote{http://weaviate.io.}, ZillizCloud\footnote{http://zilliz.com/.}, Milvus\footnote{http://milvus.io.}, ElasticCloud\footnote{http://elastic.co.}, and Pinecone\footnote{http://pinecone.io.}. The comparison of VDBs includes both the attributes and characteristics of different VDBs, as well as a comparison of their loading capacity and search performance.

\subsection{The Comparison of Features and Characteristics of Vector Databases}

The characteristics of VDBs directly affect their performance in practical applications. Therefore, gaining a deep understanding of these databases' features is essential for selecting the most suitable one. 
As shown in Table~\ref{tab:FEATURES OF VECTOR DATABASES}, we compare several popular VDBs, focusing on their differences in indexing methods, query types, distance functions, scalability, maximum dimension, and support for data management features such as replication, sharding, and partitioning. 

It can be observed from Table ~\ref{tab:FEATURES OF VECTOR DATABASES} that all VDBs support NNS and ANNS.
However, the implementation strategies and optimizations for these searches vary significantly across databases, depending on their underlying indexing methods and architectural designs. For example, the indexing methods and distance functions are not exactly the same across databases, but there are commonalities. For instance, all databases except Pinecone support graph-based methods, which indicates that graph-based methods are widely adopted for their ability to handle complex relationships and data structures. Additionally, the majority of databases also support three distance functions: inner product, cosine similarity, and Euclidean distance. For details on the indexing methods and distance functions supported by different databases, see Table~\ref{tab:Overview of Supported Distance Functions in Vector Databases} and Table~\ref{tab:Overview of Supported Indexing Methods in Vector Databases} below. 
\begin{table*}[thbp]
		\fontsize{6}{8}\selectfont    
		\centering
\vspace{-2mm}
		\caption{FEATURES OF VECTOR DATABASES}
\vspace{-1mm}
\label{tab:FEATURES OF VECTOR DATABASES}
		\setlength{\tabcolsep}{0.9mm}
            \begin{threeparttable}
		\begin{tabular}{ccccccccccccccc}
			\toprule[1pt] 
                Database  & \multicolumn{2}{c}{Query Types} &  \multicolumn{5}{c}{Indexing  Methods} & NSD  & \multicolumn{2}{c}{Scalability }  &  Replication & Sharding & Partitioning & \makecell[c]{Maximum  \\  Dimension} \\
                \cmidrule(lr){2-3} \cmidrule(lr){10-11} \cmidrule(lr){4-8} 
				   & ANNS & NNS & \makecell[c]{Brute\\ Force} & \makecell[c]{Tree\\ Based}  &  \makecell[c]{Hash\\ Based}   & \makecell[c]{Graph\\ Based} & \makecell[c]{Quantization\\ Based} & & \makecell[c]{Horizontal \\ Scaling} & \makecell[c]{Vertical \\ Scaling} 
                   &  &  &  &   \\
                \midrule
				PgVector  &\large $\checkmark$  &\large $\checkmark$ &\large $\checkmark$  &\large $\checkmark$   &\large  $\checkmark$ &\large  $\checkmark$ &\large  $\checkmark$
                & 7 &\large $\checkmark$ &\large $\checkmark$ 
                &\large $\checkmark$ &\large $\checkmark$ &\large$\checkmark$ & 16,000\\
                QdrantCloud &\large $\checkmark$  &\large $\checkmark$ &\large $\checkmark$  &\large $\times$ &\large $\times$ &\large $\checkmark$ &\large $\checkmark$ & 4 &\large $\checkmark$ &\large $\checkmark$ 
                &\large $\checkmark$ &\large $\checkmark$ &\large $\checkmark$ & 65,535\\ 
                WeaviateCloud
                  &\large $\checkmark$  &\large $\checkmark$ &\large $\checkmark$  &\large $\times$ &\large $\times$ &\large $\checkmark$
                &\large $\times$  & 6 &\large $\checkmark$   &\large $\times$ 
                &\large $\checkmark$ &\large $\checkmark$ &\large $\checkmark$  & 65,535\\
                 ZillizCloud &\large $\checkmark$  &\large $\checkmark$ &\large  $\checkmark$   &\large $\times$  &\large $\checkmark$ &\large $\checkmark$  &\large $\checkmark$ & 4
                &\large $\times$ &\large $\checkmark$ 
                &\large $\checkmark$ &\large $\checkmark$ &\large $\checkmark$ & 32,768\\
                   Milvus &\large $\checkmark$  &\large $\checkmark$ &\large $\checkmark$ &\large $\times$  &\large $\times$ &\large $\checkmark$  &\large $\checkmark$ & 6
                &\large $\checkmark$ &\large $\times$ 
                &\large $\checkmark$ &\large $\checkmark$ &\large $\checkmark$ & 32,768\\
                   ElasticCloud  &\large $\checkmark$  &\large $\checkmark$ &\large $\checkmark$   &\large $\times$ &\large $\times$  &\large $\checkmark$    &\large $\times$  & 4
                &\large $\checkmark$ &\large $\times$ 
                &\large $\checkmark$ &\large $\checkmark$ &\large $\checkmark$ &N/A\\
                Pinecone  &\large $\checkmark$  &\large $\checkmark$ & \makecell[c]{N/A}  & \makecell[c]{N/A} & \makecell[c]{N/A}& \makecell[c]{N/A} & \makecell[c]{N/A} & 3
                &\large $\checkmark$ &\large $\checkmark$ 
                &\large $\checkmark$ &\large $\checkmark$ &\large $\checkmark$ &N/A\\
				\bottomrule[1pt]
			\end{tabular}
               \begin{tablenotes}
                \tiny
                \item[] Abbreviations: NSD \ Number of Supported Distance Functions,\ N/A \  Unknown,\ $\checkmark$ \  Support,\ $\times$ \  Not\;Support
               \item[] The database information listed above is based on data up to December 1, 2024.
                \end{tablenotes}
                \end{threeparttable}
                \vspace{-3mm}
    \end{table*}
    
\begin{table}[thbp]
        \fontsize{5.2pt}{6pt}\selectfont
        \setlength\tabcolsep{0pt}
        \renewcommand{\baselinestretch}{0.9}
		\centering
        
\vspace{-3mm}
		\caption{Overview of Supported Distance Functions in VDBs}
\vspace{-1mm}
		\label{tab:Overview of Supported Distance Functions in Vector Databases}
		\setlength{\tabcolsep}{0.7mm}
            \begin{threeparttable}
		\begin{tabular}{c|cccccccc}
		\toprule[1pt] 
            \multicolumn{2}{c}{} & PgVector & QdrantCloud & WeaviateCloud &ZillizCloud &Milvus & ElasticCloud &Pinecone  \\
            \multirow{2}{*}{\rotatebox{90}{ Distance Function  \ \ \ \ \ \ \ \ }} & Inner Product & \large $\checkmark$  & \large $\checkmark$  & \large $\checkmark$  & \large $\checkmark$  & \large $\checkmark$  & \large $\checkmark$  &\large $\checkmark$ \\ 
           & Cosine Similarity & \large $\checkmark$  & \large $\checkmark$  & \large $\checkmark$  & \large $\checkmark$  & \large $\checkmark$  & \large $\checkmark$  &\large $\checkmark$ \\
           & Manhattan Distance  & \large $\checkmark$  & \large $\checkmark$  & \large $\checkmark$  & \large $\times$  & \large $\times$  & \large $\times$  &\large $\times$\\
           & Hamming Distance  & \large $\checkmark$  & \large $\times$  & \large $\checkmark$  & \large $\checkmark$  & \large $\checkmark$  & \large $\times$  &\large $\times$\\
           & Jaccard Distance  & \large $\checkmark$  & \large $\times$  & \large $\times$  & \large $\checkmark$  & \large $\checkmark$  & \large $\times$  &\large $\times$\\
           & Taxicab Distance  & \large $\checkmark$  & \large $\times$  & \large $\times$  & \large $\times$  & \large $\times$  & \large $\times$  &\large $\times$\\
           & Euclidean Distance  & \large $\checkmark$  & \large $\checkmark$  & \large $\checkmark$  & \large $\times$  & \large $\checkmark$  & \large $\checkmark$  &\large $\checkmark$\\
           & Structural Similarity & \large $\times$  & \large $\times$  & \large $\times$  & \large $\times$  & \large $\checkmark$  & \large $\times$  &\large $\times$\\
           & Max Inner Product & \large $\times$  & \large $\times$  & \large $\times$  & \large $\times$  & \large $\times$  & \large $\checkmark$  &\large $\times$\\
           
            \end{tabular}
            \begin{tablenotes}
                \tiny
                \item[] Abbreviations:  $\checkmark$ \  Support,\ $\times$ \  Not\;Support
               \item[] The database information listed above is based on data up to December 1, 2024.
                \end{tablenotes}   
            \end{threeparttable}
\end{table}   

\begin{table}[thbp]
        \fontsize{5.2pt}{6pt}\selectfont
        \setlength\tabcolsep{0.1pt}
        \renewcommand{\baselinestretch}{0.9}
        
		\centering
\vspace{-3mm}
		\caption{Overview of Supported Indexing Methods in VDBs}
\vspace{-1mm}
		\label{tab:Overview of Supported Indexing Methods in Vector Databases}
		\setlength{\tabcolsep}{0.5mm}
            \begin{threeparttable}
		\begin{tabular}{c|cccccccc}
		\toprule[1pt] 
            \multicolumn{2}{c}{} & PgVector & QdrantCloud & WeaviateCloud &ZillizCloud &Milvus & ElasticCloud &Pinecone  \\
            \multirow{2}{*}{\rotatebox{90}{ Indexing Method  \ \ \ \ \ \ \ \ \ \ \ \ \ \ \ \ \ \ \ \ \ \ \ \ \ \ \ \  \ \ \ }} & HNSW & \large $\checkmark$  & \large $\checkmark$  & \large $\checkmark$  & \large $\checkmark$  & \large $\checkmark$  & \large $\checkmark$  & N/A \\ 
           & Flat & \large $\times$  & \large $\times$  & \large $\checkmark$  & \large $\times$  & \large $\checkmark$  & \large $\times$  & N/A \\
           & BINFlat  & \large $\times$  & \large $\times$  & \large $\times$  & \large $\times$  & \large $\checkmark$  & \large $\times$  & N/A\\
           & IVF\_Flat  & \large $\times$  & \large $\times$  & \large $\times$  & \large $\times$  & \large $\checkmark$  & \large $\times$  & N/A\\
           & BIN\_IVF\_Flat  & \large $\times$  & \large $\times$  & \large $\times$  & \large $\times$  & \large $\checkmark$  & \large $\times$  & N/A\\
           & IVF\_SQ8  & \large $\times$  & \large $\times$  & \large $\times$  & \large $\times$  & \large $\checkmark$  & \large $\times$  & N/A\\
           & IVF\_PQ  & \large $\times$  & \large $\times$  & \large $\times$  & \large $\times$  & \large $\checkmark$  & \large $\times$  & N/A\\
           & B-tree & \large $\checkmark$  & \large $\times$  & \large $\times$  & \large $\times$  & \large $\times$  & \large $\times$  & N/A\\
           & LSH & \large $\times$  & \large $\times$  & \large $\times$  & \large $\checkmark$  & \large $\times$  & \large $\times$  & N/A\\
           & BRIN & \large $\checkmark$  & \large $\times$  & \large $\times$  & \large $\times$  & \large $\times$  & \large $\times$  & N/A\\
           & Inverted\_File\_Index & \large $\checkmark$  & \large $\times$  & \large $\times$  & \large $\times$  & \large $\checkmark$  & \large $\checkmark$  & N/A\\
           & SPARSE Inverted Index  & \large $\times$  & \large $\times$  & \large $\times$  & \large $\times$  & \large $\times$  & \large $\times$  &  N/A\\
           & SPARSE WAND  & \large $\times$  & \large $\times$  & \large $\times$  & \large $\times$  & \large $\checkmark$  & \large $\times$  & N/A\\
           & GIST & \large $\checkmark$  & \large $\times$  & \large $\times$  & \large $\times$  & \large $\times$  & \large $\times$  & N/A\\
           & GIN & \large $\checkmark$  & \large $\times$  & \large $\times$  & \large $\times$  & \large $\times$  & \large $\times$  & N/A\\
           & DiskANN & \large $\times$  & \large $\checkmark$  & \large $\times$  & \large $\checkmark$  & \large $\times$  & \large $\times$  & N/A\\
           & SCANN  & \large $\times$  & \large $\times$  & \large $\times$  & \large $\checkmark$  & \large $\times$  & \large $\times$  & N/A\\
        & Sparse Vector Index  & \large $\times$  & \large $\checkmark$  & \large $\times$  & \large $\times$  & \large $\times$  & \large $\times$  & N/A\\
           & Parameterized index & \large $\times$  & \large $\checkmark$  & \large $\times$  & \large $\times$  & \large $\times$  & \large $\times$  & N/A \\
            
            \end{tabular}
            \begin{tablenotes}
                \tiny
                \item[] Abbreviations:\ N/A \  Unknown,\ $\checkmark$ \  Support,\ $\times$ \  Not\;Support
               \item[] The database information listed above is based on data up to December 1, 2024.
                \end{tablenotes}   
            \end{threeparttable}
\vspace{-3mm}
            
\end{table}   
Scalability is a critical factor in evaluating the performance and flexibility of VDBs, especially for large-scale and high-demand applications. Scalability is typically categorized into horizontal scaling and vertical scaling. 
Horizontal scaling refers to a database's ability to distribute data and computation across multiple nodes, allowing it to handle large datasets and high query throughput. This approach is particularly beneficial for cloud-native environments and distributed architectures, where data is sharded and replicated across multiple machines. In contrast, vertical scaling involves upgrading a single machine with more resources, such as additional CPU power or memory, to manage increased workloads. Both scaling methods offer distinct advantages depending on the application’s requirements and the environment in which the database operates. Specifically, PgVector, QdrantCloud, and Pinecone support both horizontal and vertical scaling modes, while WeaviateCloud, Milvus, and ElasticCloud only support horizontal scaling. ZillizCloud is the only one that supports only vertical scaling. Although the level of support for scalability varies across databases, most exhibit strong capabilities in data storage and backup. Specifically, all databases, except for ElasticCloud, for which no relevant information was found, support Replication, Sharding, and Partitioning. These features ensure fault tolerance, efficient data distribution, and flexible query handling.

The last column of table~\ref{tab:FEATURES OF VECTOR DATABASES} provides statistics on the maximum vector dimensions supported by each database. It can be observed that, with the exception of ElasticCloud and Pinecone, for which no relevant information was available, most of the listed VDBs support a total vector dimension in the range of tens of thousands, with the maximum supported dimensions ranging from 16,000 to 65,535. It should be noted that QdrantCloud has default support for up to 65,535 dimensions, though this can be configured to support higher dimensions. 

\subsection{The Comparison of Loading Capacity and Search Performance of VDB}

In this subsection, we have opted to use the performance results obtained from the existing benchmarking tool, VectorDBBench(A Benchmark Tool for VectorDB)\footnote{https://github.com/zilliztech/VectorDBBench?tab=readme-ov-file}, rather than conducting our own tests. This decision is based on the tool's comprehensive and standardized testing methodology, which provides reliable, reproducible results across various VDBs. By utilizing pre-existing data, we ensure consistency and comparability, as these results have been generated under controlled conditions, following established benchmarks. 

VectorDBBench provides a comprehensive performance analysis by evaluating VDBs based on metrics such as Queries Per Second (QPS), recall rate, latency (the time required for each query from submission to system response), load duration, and maximum load count (The maximum number of vectors a database can successfully insert or store in a single loading operation). Its testing methodology employs a relative scoring mechanism to ensure fair comparisons. For QPS, the highest observed value among all tested databases serves as the reference baseline; for latency, the lowest observed value among all tested databases is used as the baseline, with an additional 10 ms adjustment to avoid distortions when latency is very low. For systems that fail or encounter timeouts in a specific test case, their scores are penalized by assigning a value proportionally worse than the lowest-performing result, using a factor of two. For example, in the case of QPS, the score is reduced to half of the minimum observed value, while for latency, it is increased to twice the maximum observed value. 
The formulas for calculating QPS and latency metrics for VDB $x$ are as follows: 

\begin{equation}
\label{eq:QPS}
QPS_{x}=\frac{origin\_QPS_{x}}{base\_QPS} \times 100
\end{equation}

\begin{equation}
\label{eq:Latency}
Latency_{x}=\frac{base\_Latency + 10ms}{origin\_Latency_{x} + 10ms} \times 100
\end{equation}
where $origin\_QPS_{x}$ and $origin\_Latency_{x}$ represent the original QPS value and original latency value, respectively, measured for database $x$ during the test. $base\_QPS$ and $base\_Latency$ is the reference baseline. 

\begin{table}[thbp]
        \fontsize{5.2pt}{6pt}\selectfont
        \setlength\tabcolsep{0.8pt}
        \centering
\vspace{-3mm}
        \caption{VECTOR DATABASE EVALUATION TEST CASES}
\label{tab:VECTOR DATABASE EVALUATION TEST CASES}
        \setlength{\tabcolsep}{0.7mm}
        \begin{threeparttable} 
        \begin{tabular}{ccccccc}
            \toprule[1pt]
           Case No.  & Case Type & Dataset &Dataset Size & Vector Dimensions & Filtering Rate  & Test Metrics   \\ 
           \midrule
            1 &Capacity  & \makecell[c]{SIFT\tnote{1}} & 500K & 128& \makecell[c]{N/A} & NIV \\
            \specialrule{0em}{2pt}{4pt}
            2 &Capacity  & \makecell[c]{GIST\tnote{2} } &  100K &960  &\makecell[c]{N/A} &NIV \\
            \specialrule{0em}{2pt}{4pt}
            3 &Search Performance  & \makecell[c]{Google C4\tnote{3}} & 500K & 1536 &\makecell[c]{N/A} & IBT, R, L, MQPS  \\
            \specialrule{0em}{2pt}{4pt}           
            \bottomrule
        \end{tabular}
        \begin{tablenotes}
                \tiny
                \item[1] http://corpus-texmex.irisa.fr/
                \item[2] http://corpus-texmex.irisa.fr/
                \item[3] The\;processed\;version\;of\;Google\;C4 dataset (https://huggingface.co/datasets/allenai/c4)
                \item[] Abbreviations: N/A. Not\;Applicable, NIV.   Number\;of\;inserted\;vector, IBT. Index\;building\;time, R. Recall, L. Latency, MQPS. Maxiumum\;QPS\ 
               
                \end{tablenotes}
                \end{threeparttable}  
\end{table}

Specifically, as shown in Table~\ref{tab:VECTOR DATABASE EVALUATION TEST CASES}, the VDB evaluation consists of a series of test cases designed to assess capacity, search performance, and filtering search performance. Capacity cases (Cases 1 and 2) measure the database's ability to handle large datasets, focusing on the number of inserted vectors using SIFT and GIST datasets. Search performance cases (Case 3) evaluate index building time, recall, latency, and maximum QPS using the Google C4 dataset.

\begin{figure}[]
	\centering
    \vspace{-3mm}
	\includegraphics[width=3in]{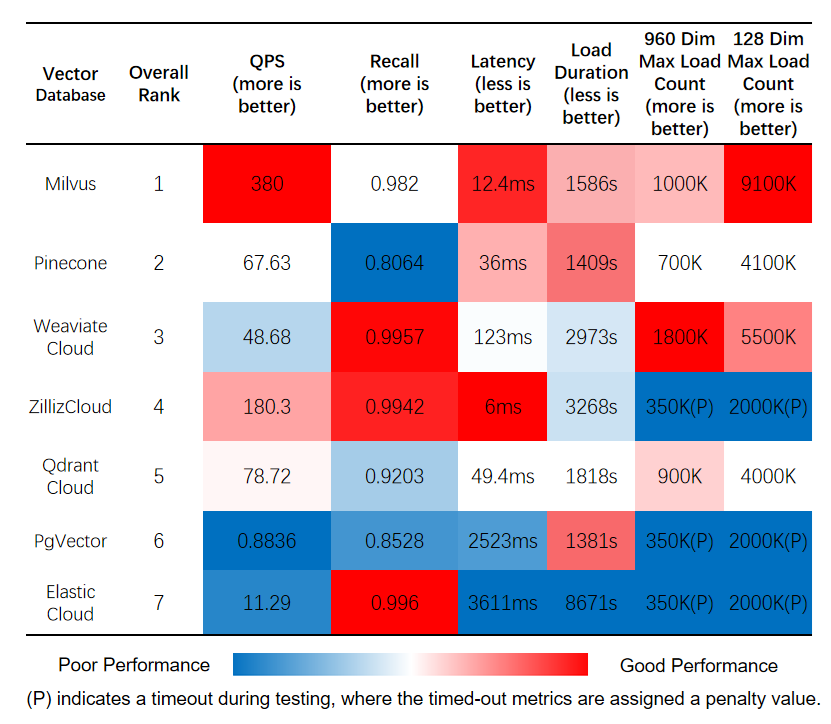}
    \vspace{-3mm}
	\caption{Performance Test Results for VDBs}
    \vspace{-3mm}
	\label{fig:Performance Test Results for Vector Databases}
\end{figure}
The VDB versions involved in the performance tests are as follows: Milvus-2c8g-hnsw-v2.2.12 (hereafter referred to as Milvus), Pinecone-p1.x1 (hereafter referred to as Pinecone), WeaviateCloud-standard (hereafter referred to as Weaviate Cloud), ZillizCloud-2cu-cap-v2023.6 (hereafter referred to as ZillizCloud), QdrantCloud-2c8g-1node (hereafter referred to as QdrantCloud), PgVector-2c8g (hereafter referred to as PgVector), and ElasticCloud-upTo2.5c8g (hereafter referred to as ElasticCloud). To ensure minimal differences in hardware performance across the tested databases, a configuration of 2 CPUs and 8GB of memory was specifically selected. For VDBs that do not meet this hardware requirement, similar configurations were chosen as closely as possible. The specific test results are shown in the Figure~\ref{fig:Performance Test Results for Vector Databases}. The overall ranking in Figure~\ref{fig:Performance Test Results for Vector Databases} is calculated by averaging the rankings of each sub-test item, with the final overall ranking determined in ascending order of the average values. According to the overall ranking, Milvus ranks first with its comprehensive performance, achieving a QPS of 380, latency of 12.4 milliseconds, and outstanding performance in load capacity. In contrast, ElasticCloud ranks last, with a QPS of 11.29 and latency as high as 361 milliseconds. In terms of recall rate, all databases perform close to 1.0, indicating little difference in query accuracy. ZillizCloud demonstrates the best latency performance, at only 6 milliseconds, but its load capacity is relatively low. It is worth noting that ZillizCloud, ElasticCloud, and PgVector were assigned penalty values (P) due to timeouts in the load capacity tests, which may have been caused by network issues and should not be taken as a definitive measure of their actual performance. Overall, no single database ranked in the top three across all tests, indicating that different databases may have their own strengths. Additionally, the overall rank is simply a straightforward average of the rankings in each test item, without applying weighted averages based on specific tasks, so it should be regarded as a preliminary reference. 

A deeper analysis of these performance results reveals that multiple factors influence the outcomes, including hardware configuration, deployment mode, index parameter settings, and data distribution. In this benchmark, all databases were configured with 2 CPU cores and 8 GB of memory to minimize hardware discrepancies. However, different databases exhibit varying efficiency in resource utilization. For instance, ZillizCloud demonstrated outstanding low-latency performance but faced limitations in load capacity due to network timeouts and indexing strategies; Milvus excelled in both QPS and load capacity, likely due to its efficient index implementation and parallel processing capabilities. It is important to note the typical trade-off between QPS and recall: higher query throughput often comes at the cost of lower recall. In this evaluation, Pinecone achieved a relatively low recall (0.8064) but still maintained competitive QPS, suggesting that it may prioritize speed over accuracy in certain scenarios. The timeouts observed in PgVector and ElasticCloud during load tests may be attributed to data ingestion methods, network conditions, or index construction efficiency, and should not be interpreted as definitive indicators of their performance. Therefore, when selecting a VDB, users should consider their specific application requirements—such as real-time responsiveness, data scale, and hardware constraints—and are encouraged to conduct customized evaluations in their own environments to validate performance.

Beyond the established vector databases evaluated above, recent research has introduced HARMONY, a distributed ANNS system that addresses load imbalance and high communication overhead caused by traditional partition strategies\cite{xu2025harmony}. It uses multi-granularity partitioning to balance load and minimize communication, plus an early-stop pruning mechanism to reduce overhead. Experiments show HARMONY achieves 4.63× higher throughput than leading distributed vector databases and 58\% improvement for skewed workloads, offering a promising direction for future VDB development.

\vspace{-3mm}
\section{Challenges}

\subsection{High-Dimensional Vector Index and Search}
VDBs require efficient indexing and searching of billions of vectors in hundreds or thousands of dimensions, which poses a huge computational and storage challenge. Traditional indexing methods, such as B-trees or hash tables, are not suitable for high-dimensional vectors because they suffer from dimensionality catastrophe. Therefore, VDBs need to use specialized techniques such as ANN search, hashing, quantization, or graph-based search to reduce complexity and improve the accuracy of vector similarity search.

\vspace{-3mm}
\subsection{Support for Heterogeneous Vector Data Types}
VDBs need to support different types of vector data, such as dense vectors, sparse vectors, binary vectors, and so on. Each type of vector data may have different characteristics and requirements, such as dimensionality, sparsity, distribution, similarity metrics, and so on. Therefore, VDBs need to provide a flexible and adaptive indexing system to handle various vector data types and optimize their performance and availability.

\vspace{-3mm}
\subsection{Distributed Parallel Processing Support}
VDBs need to be scalable to handle large-scale vector data and queries that may exceed the capacity of a single machine. Therefore, VDBs need to support distributed parallel processing of vector data and queries across multiple computers or clusters. This involves challenges such as data partitioning, load balancing, fault tolerance, and consistency.

\vspace{-3mm}
\subsection{Integration with Emerging Application Scenarios}
Currently, many emerging application scenarios remain underexplored. For instance, the incremental k-NN search~\cite{zhao2008incremental} adopted by recommendation and e-commerce platforms faces significant challenges due to the vast imbalance between the volume of vector data processed and the data displayed to users. This method cannot be effectively supported by most VDBs~\cite{pan2024survey}. Furthermore, with the latest advancements in sparse vector technology, integrating these technologies into VDBs to enable hybrid retrieval (combining keyword and vector retrieval methods) is increasingly regarded as a best practice. Such hybrid systems must manage large datasets while enhancing computational efficiency and maintaining retrieval quality.

\vspace{-3mm}
\subsection{Data Security and Privacy Protection}
As cyber threats intensify and regulatory requirements become more stringent, data security and privacy protection for databases have become top priorities~\cite{farayola2024data}. Compared to the comprehensive data security and privacy protection features of traditional relational databases, VDBs are still in the early stages. Unlike traditional relational databases, VDBs typically handle large amounts of high-dimensional embedding vectors, which pose higher risks of privacy breaches during storage, querying, and transmission. This is especially true in cloud platforms, where data is often stored in an unencrypted form or transmitted between different nodes, increasing the potential attack surface~\cite{asaad2024enhancing, amaithi2024systematic}. In the future, VDBs will need to not only leverage traditional data security and privacy protection methods to build a robust security framework but also integrate emerging technologies, such as blockchain tables~\cite{s23167172} (to ensure data immutability) and AI-based anomaly detection (for proactive threat management), to adapt to specific application scenarios.

\vspace{-3mm}
\section{Synergy of LLMs and VDBs}

When processing natural language, Large Language Models (LLMs) need to convert natural language into high-dimensional vectors, which requires robust capabilities for storing and retrieving high-dimensional vectors. Meanwhile, the inherent issues of LLMs, such as "hallucinations" (generating content that seems plausible but is factually incorrect) and "forgetfulness" (difficulty in remembering or utilizing early information), also necessitate supporting facilities like vast external knowledge bases to mitigate them, all of which cannot be achieved without the support of VDBs. Therefore, the integration of LLMs and VDBs is an inevitable trend. Among the approaches, integrating VDBs into LLM systems is a promising method. In the following section, we will delve into the fusion, mutual influence, and potential applications of the two, providing references for scientific research and industrial applications.

\vspace{-3mm}
\subsection{VDBs for LLMs}

\begin{figure}[thbp] 
    \centering
    \vspace{-3mm}
    \includegraphics[width=0.5\textwidth]{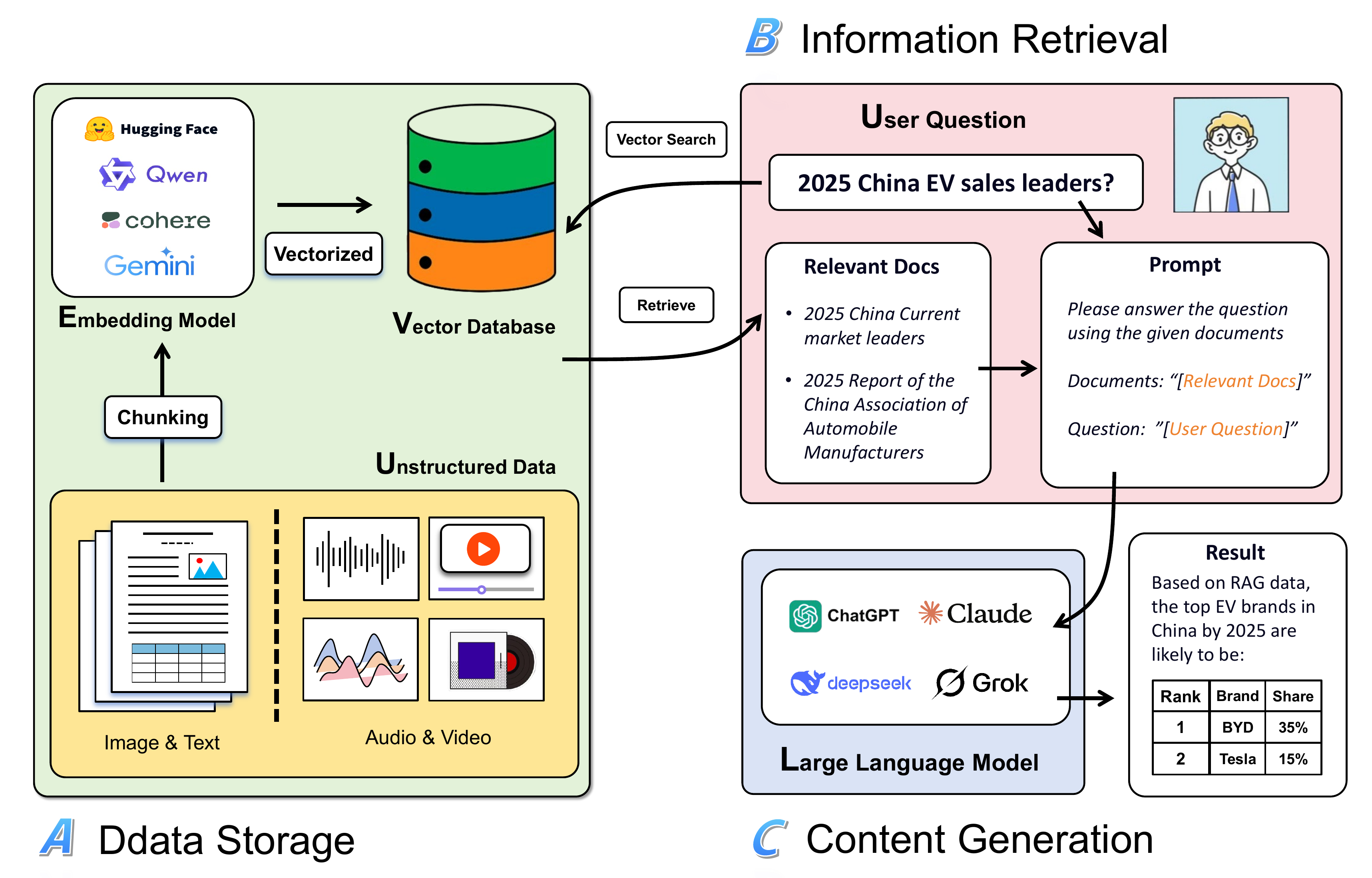} 
    \vspace{-3mm}
    \caption{A common workflow of RAG}
    \vspace{-3mm}
    
    \label{fig:A common workflow of RAG}
\end{figure}

LLMs are characterized by large model capacity and vast data corpus\cite{openai2024gpt}. With hundreds of billions (or more) of parameters and extensive textual training, they are highly adept at comprehending human knowledge and instructions~\cite{shanahan2023talking}. However, LLMs do have certain shortcomings, though~\cite{zhao2023survey}. One major shortcoming is hallucinations, where the model generates a response that is factually inaccurate. This shortcoming is mainly caused by the following issues, including knowledge limitations confined by the training corpus; the internal knowledge in LLMs cannot be updated, resulting in outdated knowledge; and LLMs may also introduce systematic errors due to the large dataset used for training. Another shortcoming is the oblivion problem. LLMs have been found to have the inclination to forget the previous input information and also exhibit catastrophic forgetting behavior. 
In response to these issues, VDBs can offer robust support for LLMs in the following aspects:

\textbf{1) VDBs as an External Knowledge Base: Retrieval-Augmented Generation (RAG).} 
VDBs as External Knowledge Base: Retrieval-Augmented Generation (RAG). Retrieval-Augmented Generation (RAG) technology is an innovation in the field of artificial intelligence that integrates information retrieval with language generation models~\cite{lewis2020retrieval,lewis2021retrievalaugmentedgenerationknowledgeintensivenlp}. This technology significantly enhances the performance of Large Language Models (LLMs) in knowledge-intensive tasks such as question answering, text summarization, and content generation by retrieving relevant information from an external knowledge base and inputting it as a prompt to LLMs. RAG encompasses core processes including retrieval, generation, and augmentation. Figure~\ref{fig:A common workflow of RAG} illustrates a typical workflow of RAG when integrated with LLMs. The system's operational workflow mainly consists of three core components: data storage, information retrieval, and content generation.

The RAG workflow begins with the data storage phase. During this phase, externally collected unstructured data (text, images, audio, and video) undergoes preprocessing, is divided into smaller chunks, and is converted into vectors through an embedding model to capture semantic representations. These vectors are then stored in a VDB for subsequent vector retrieval.

Next is the information retrieval phase. When a user poses a question in the form of a prompt, the embedding model generates a vector for the query and retrieves the most semantically similar data chunks from the VDB. The retrieved results are converted back from vector format to their original format and returned to the user.

Finally, in the content generation phase, the Large Language Model (LLM) integrates the user's original question and the retrieved information, selects an appropriate prompt template based on the task type, processes the prompt, and generates the final answer.

\textbf{2) VDBs as a Cost-effective Semantic Cache.} The operation of LLMs requires substantial computational resources, and frequent API calls to third-party models incur high costs. By integrating VDBs, they can be utilized as a semantic cache to achieve efficient semantic matching through query embeddings. This method, by storing query embeddings, can reduce redundant API calls, lower costs, and improve response speed and efficiency~\cite{regmi2024gpt}. Moreover, the combination of VDBs and LLMs demonstrates strong scalability and adaptability, enabling efficient handling of a large volume of queries, stable performance under fluctuating workloads, and support for multiple embedding models and configurations to meet diverse deployment needs ~\cite{bang2023gptcache}. Using VDBs as a semantic cache for GPT is a feasible strategy to promote the large-scale application of LLMs.

\textbf{3) VDBs as a Reliable Memory of LLMs.} A notable shortcoming of current LLMs is their lack of strong long-term memory capabilities~\cite{hatalis2023memory}. Memory systems can enhance the intelligence of LLMs, enabling them to possess autonomous capabilities and improve their performance in various tasks. VDBs can act as a memory tool for LLMs, supporting the storage of historical information and allowing LLMs to effectively store different types of historical interaction information, such as knowledge, dialogue, and related task information. Additionally, LLMs lack the ability to dynamically update knowledge and engage in few-shot learning. VDBs can continuously update new information to ensure that responses are made based on the latest and most relevant data.

\vspace{-3mm}
\subsection{LLMs for VDBs}

In addition, LLMs in turn can empower databases. AI technology has been proved to perform well in many data management tasks, such as data processing, database optimization, and data analysis. However, traditional machine learning algorithms are unable to solve generalization and inference problems. For example, traditional machine learning algorithms have difficulty in adapting to different databases, different query workloads, and different hardware environments, making them unable to solve the generalizability and inference problems in data management tasks. In addition, traditional machine learning algorithms cannot satisfy the need for contextual understanding and multi-step reasoning in optimization scenarios such as database diagnosis, root cause analysis, etc. However, LLMs bring promising solutions to the above problems~\cite{zhou2024db}.

\textbf{LLMs assist database management tasks.} Large Language Models (LLMs) are actively contributing to database management and driving innovation in the field of data management. For instance, LLMs can analyze abnormal database metrics, report the root causes and potential solutions to administrators, and also serve as a natural language interface to convert user requests into executable queries. Moreover, after being integrated with databases, LLMs can perform exceptionally well on new tasks with minimal fine-tuning, making them more adaptable to changes in database schemas, data, and hardware. By guiding model inference through prompts, LLMs achieve a user-friendly interface, providing an intuitive experience without the need for extensive training data or multiple iterations to incorporate user feedback. Additionally, LLMs can extract insights from database components (such as documents and code), integrating their strengths to enhance database performance and compensate for the weaknesses of individual components~\cite{li2024llm}.

\textbf{LLMs smarten vector data handling}. The deep integration of LLMs with VDBs has pioneered innovative application scenarios for data-driven workflows, encompassing content generation, knowledge enhancement, and system optimization. By combining semantic understanding with vectorized retrieval, LLMs can generate customized texts (e.g., thematic articles, stylized summaries) based on vector inputs, enrich ambiguous texts with additional details (e.g., supplementing statistical data or case studies), and facilitate cross-language, cross-domain text transformations (e.g., multilingual simplification of legal documents). Furthermore, LLMs significantly optimize the management tasks of VDBs: they recommend configuration parameters by analyzing historical performance data to improve system stability, automatically diagnose performance bottlenecks while generating interpretable reports, and efficiently process heterogeneous data through semantic analysis (e.g., schema matching and error correction). These applications not only reduce the cost of manual intervention but also extend the generalization capabilities of traditional methods through adaptive solutions, highlighting the core value of LLMs in enhancing the intelligence and scalability of VDBs~\cite{tang2023does, albalak2024survey, fan2024cost,zhou2023d,huang2024llmtune, chang2023prompt, whitehouse2023llmpowered}.

\begin{figure}[t]
    \centering
    \vspace{-3mm}\includegraphics[width=1.0\linewidth, keepaspectratio]{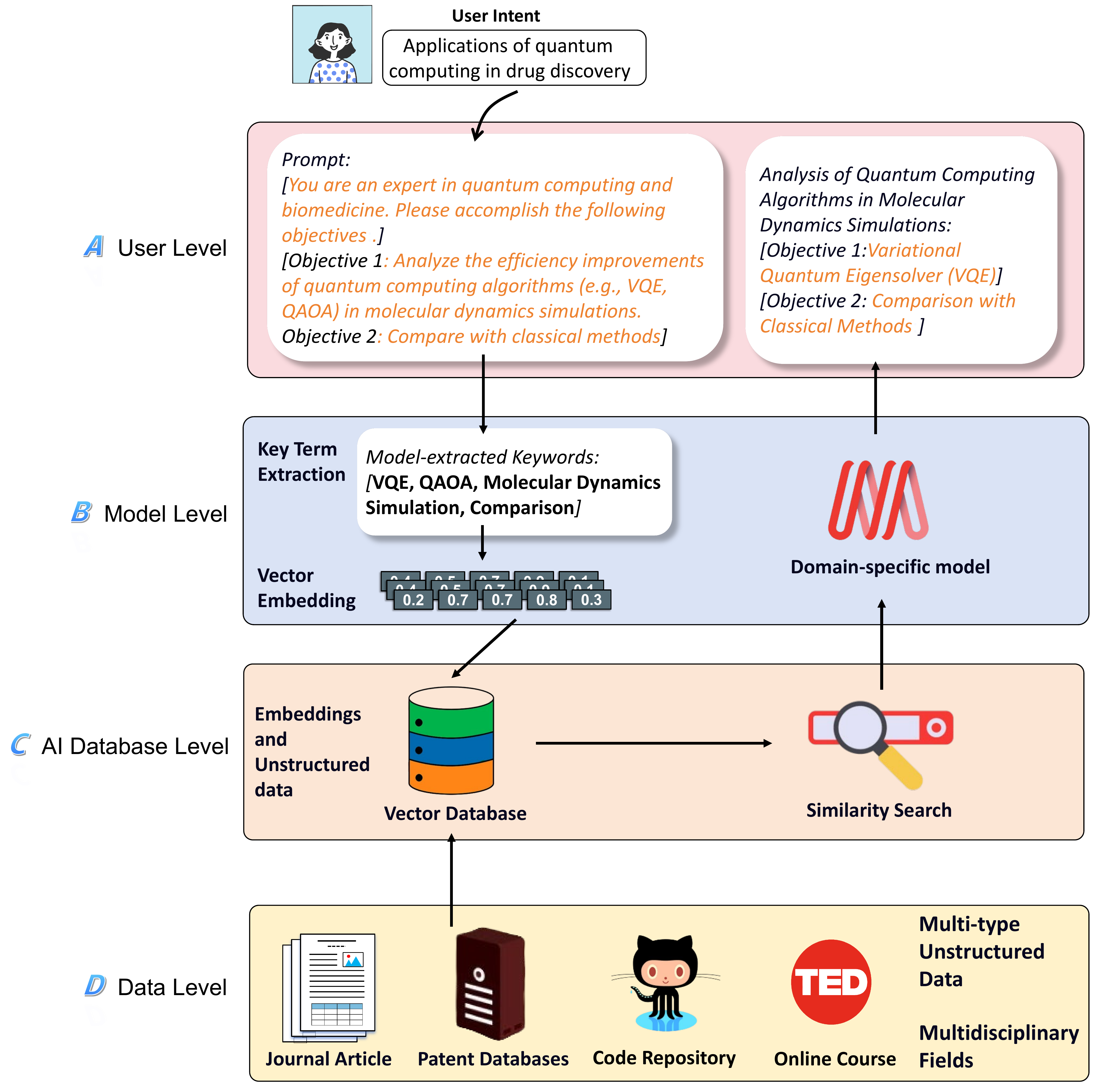} 
    \vspace{-3mm}
    \caption{A common workflow of Retrieval-Augmented Generation (RAG).}
    \vspace{-3mm}
    \label{fig:rag_workflow} 
\end{figure}

\vspace{-3mm}
\subsection{A General LLMs and VDBs Synergized Framework}

For a framework that incorporates a large language model
and a VDB, as shown in Figure~\ref{fig:A common workflow of RAG}, can be understood by splitting it into four levels: the user level, the model level, the AI database
level, and the data level, respectively.
For a user who has never been exposed to large language
modeling, it is possible to enter natural language to describe
their problem. For a user who is proficient in large language
modeling, a well-designed prompt can be entered.
The LLM next processes the problem to extract the key-
words in it, or in the case of open-source LLMs, the corre-
sponding vector embeddings can be obtained directly.
The VDB stores unstructured data and their joint
embeddings. The next step is to go to the VDB to
find similar nearest neighbors. The ones obtained from the
sequences in the big language model are compared with the
vector encodings in the VDB by means of the NNS
or ANNS algorithms. And different results are derived through
a predefined serialization chain, which plays the role of a
search engine.
If it is not a generalized question, the results derived need to
be further put into the domain model; for example, imagine we
are seeking an intelligent scientific assistant, which can be put
into the model of AI4S to get professional results. Eventually
it can be placed again into the LLM to get coherent generated
results.
For the data layer located at the bottom, one can choose
from a variety of file formats such as PDF, CSV, MD, DOC,
PNG, SQL, etc., and its sources can be journals, conferences,
textbooks, and so on. Corresponding disciplines can be art,
science, engineering, business, medicine, law, etc.

\vspace{-4mm}
\section{Conclusion}

In this paper, we provide a comprehensive and up-to-date literature review on VDBs, including the key algorithms, storage, and retrieval methods. 
We also compare representative VDB systems, analyze their design trade-offs, and discuss their strengths, limitations, and typical use cases. 
Furthermore, we identify key challenges and outline potential research directions, including improved indexing and closer integration with LLMs.  
We believe this survey offers a solid reference for researchers and practitioners, and contributes to a clearer understanding of the current state and future direction of VDBs. 

\bibliography{references}
\printcredits

\end{document}